\documentclass[preprint,3p,12pt]{elsarticle}

\usepackage{mathrsfs}
\usepackage{amsmath}
\usepackage{stmaryrd}
\usepackage{bbding}
\usepackage{dcolumn}
\usepackage{graphicx}
\usepackage{amsfonts}
\usepackage{amssymb}
\usepackage{psfrag}
\usepackage{wrapfig}
\usepackage{subfigure}
\usepackage{makeidx}
\usepackage{bm}
\usepackage{epsf}
\usepackage{epsfig}
\usepackage{setspace}
\usepackage{graphicx}
\usepackage{epstopdf}
\usepackage{psfrag}
\usepackage{subfigure}
\usepackage{color}
\usepackage{comment}
\usepackage{algorithm}
\usepackage{algorithmic}
\newcommand \td {\mathrm{~d}}
\journal{}

\begin{document}

\begin{frontmatter}



\title{A single-stage high-order compact gas-kinetic scheme in arbitrary Lagrangian-Eulerian formulation}

\author[HKUST1]{Yue Zhang}
\ead{yzhangnl@connect.ust.hk}	
\author[xijiao]{Xing Ji} 
\ead{jixing@xjtu.edu.cn}
\author[iapcm]{Yibing Chen}
\ead{chen_yibing@iapcm.ac.cn}
\author[HKUST1]{Fengxiang Zhao} 
\ead{fzhaoac@connect.ust.hk}
\author[HKUST1,HKUST2,HKUST3]{Kun Xu\corref{cor1}}
\ead{makxu@ust.hk}

\address[HKUST1]{Department of Mathematics, Hong Kong University of Science and Technology, Clear Water Bay, Kowloon, Hong Kong}
\address[HKUST2]{Department of Mechanical and Aerospace Engineering, Hong Kong University of Science and Technology, Clear Water Bay, Kowloon, Hong Kong}
\address[HKUST3]{Shenzhen Research Institute, Hong Kong University of Science and Technology, Shenzhen, China}
\cortext[cor1]{Corresponding author}
\address[xijiao]{Shaanxi Key Laboratory of Environment and Control for Flight Vehicle, State Key Laboratory for Strength and Vibration of Mechanical Structures, School of Aerospace Engineering, Xi'an Jiaotong University, Xi'an, China}

\address[iapcm]{Institute of Applied Physics and Computational Mathematics, Beijing, PR China}

\begin{abstract}
This study presents the development of a compact gas-kinetic scheme using an arbitrary Lagrangian-Eulerian (ALE) formulation for structured meshes. Unlike the Eulerian formulation, the ALE approach effectively tracks flow discontinuities, such as shock waves and contact discontinuities. However, mesh motion alters the geometry and increases computational costs. To address this, two key strategies were introduced to reduce costs and enhance accuracy.
The first strategy is to use the gas-kinetic scheme to construct a third-order gas-kinetic flux, rather than the Runge-Kutta method to achieve high-order time accuracy, which allows a single reconstruction and flux calculation per time step. This approach enables direct updates of both cell-averaged flow variables and their gradients using a time-accurate flux function, facilitating compact reconstruction. Second, the significant computational expense is spent on reconstruction, which requires recalculating the reconstruction matrix at each time step due to mesh changes. A simplified fourth-order compact reconstruction using a small matrix was used to mitigate this cost. The combination of fourth-order spatial reconstruction and third-order time-accurate flux evolution ensures both high resolution and computational efficiency in the ALE framework. The tests shows that the current reconstruction is 2.4x to 3.0x faster than the previous reconstruction. Additionally,  a generalized ENO(GENO) method for handling discontinuities enhances the scheme's robustness. The numerical test cases, such as the Riemann problem, Sedov problem, Noh problem, and Saltzmann problem, demonstrated the robustness and accuracy of our method.
\end{abstract}


\end{frontmatter}


\section{Introduction}
There are two common frameworks for describing fluid motion in computational fluid dynamics: the Eulerian and the Lagrangian systems.  In the Eulerian system, the flow is described on a fixed grid, whereas in the Lagrangian system, the grid moves with the fluid velocity. While the Eulerian system is more straightforward to implement, it struggles to handle moving boundaries and to introduce dissipation when dealing with contact discontinuities. The Lagrangian system can resolve contact discontinuities well, such as material interfaces or free surfaces, but may lead to computational breakdown due to grid distortion. Therefore, the development of arbitrary Lagrangian-Eulerian (ALE) methods \cite{HIRT1974227} becomes attractive. There are two types of ALE methods: direct and indirect. Direct ALE methods involve moving both the object and the computational grid simultaneously.
In recent years, significant advancements have been made in direct ALE methods for three-dimensional flow \cite{dumbser2013arbitrary,boscheri2014direct}, particularly within the framework of the one-step ADER-WENO scheme \cite{ADER1,ADER2}. These methods utilize high-order numerical schemes to accurately capture the flow dynamics. Furthermore, researchers have developed high-order direct ALE schemes that can handle topology changes \cite{gaburro2020high}. Moreover, efforts have been made to extend the application of ALE methods to weakly compressible flows \cite{weakCom}, two-phase flows \cite{henneaux2023high}, and magnetohydrodynamics \cite{chiocchetti2021high}. These extensions enable the simulation of a broader range of physical phenomena and enhance the versatility and applicability of direct ALE methods.
Indirect ALE methods separate the motion of the object and the computational grid into three stages: the Lagrangian stage, the rezoning stage \cite{BRACKBILL1982342}, and the remapping stage \cite{KUCHARIK2014268}. During the Lagrangian stage, the flow field and grid are updated simultaneously. During the rezoning stage, a new computational grid is constructed to match the object's new position and shape, often involving techniques such as grid generation and deformation to ensure grid quality and accuracy.
In the remapping stage, the solution is transferred into the rezoned mesh.
The current research focuses on the direct ALE method with a fixed-grid topology, designed to handle shock waves and contact discontinuities. Therefore, the remapping and rezoning steps can be omitted.

In the ALE framework, the grid motion alters the geometry and increases computational costs. To achieve high-order accuracy and computational efficiency, high-order methods are necessary. However, the traditional high-order methods incur high computational cost due to the reconstruction matrix in two ways: a large reconstruction matrix that must be inverted and multiple reconstructions at each sub-time step, as in the Runge-Kutta method. The gas-kinetic scheme (GKS) is a kinetic theory-based numerical method to solve the Euler and Navier-Stokes equations \cite{xuGKS2001}. It can provide a time-accurate flux function; thus, the single-stage high-order time-integration method\cite{li2010high,luo2013high,zhou2017simplification,liu2014high} and the multi-stage multi-derivative (MSMD) method\cite{MSMD,liTwoStageFourthOrder2016,li2019two,jiThreedimensionalCompactHighorder2020} are used to update the solution with high-order temporal accuracy. 
In this work, due to the ALE framework, the grid geometry is time-dependent, requiring the reconstruction matrix to be recalculated at each time step. Therefore, a single-stage third-order gas-kinetic flux evolution model is used for this work. The third-order gas-kinetic flux was first proposed by Li et al. \cite{li2010high} and then extended by Luo et al. \cite{luo2013high} to a multidimensional scheme. Zhou et al. \cite{zhou2017simplification} further simplified the evolution model.

In addition, the gas-kinetic scheme can also provide the time-accurate flow variables at cell interfaces, and the cell-averaged gradient can be obtained via the Green-Gauss method. 
Under the initial condition of a generalized Riemann problem, a time-evolving gas distribution function is constructed in GKS to compute numerical fluxes and evaluate time-dependent flow variables at a cell interface.
As a result, both cell-averaged flow variables and their gradients can be updated directly without the need for additional equations. Compared with the DG method with different evolution equations, the cell-averaged flow variables and their gradients in GKS can be directly updated from a single gas evolution model at the cell interface.
The Compact GKS (CGKS) has been constructed on both structured and unstructured meshes in 2D and 3D cases  \cite{ji4order2018,jiThreedimensionalCompactHighorder2020,
zhaoCompactHigherorderGaskinetic2019,zhaoAcousticShockWave2020,zhaoCompactHighorderGaskinetic2022a}.
To further improve the scheme's robustness in high-speed flow simulations, additional improvements have been incorporated, such as the evolving flow variables at a cell interface with possible discontinuities \cite{zhaoDirectModelingComputational2021} and discontinuity feedback factor \cite{jiGC2021,zhang2023slidingmesh}. 
Recently, higher-order compact gas-kinetic schemes have been developed via the generalized ENO(GENO) reconstruction \cite{yang2025effective,zhang2023slidingmesh} by introducing a path function to combine the high-order linear polynomial and the low-order ENO reconstruction. This method can achieve high-order accuracy and robustness by handling discontinuities.
In recent years, high-order GKS has been extended to incorporate ALE formulation. Pan developed a high-order finite-volume ALE gas-kinetic method on structured meshes in two and three dimensions \cite{PAN2020ALE2d,PAN2021ALE3d}, which uses a non-compact WENO scheme and a two-stage fourth-order discretization.
An unstructured, compact gas-kinetic scheme has been developed \cite{zhang2024high}, which uses a third-order WENO scheme and a two-stage fourth-order discretization. 
Due to the compact stencil, a memory-reduction compact gas-kinetic scheme \cite{liu2024memory} has been developed and employed in an ALE formulation \cite{wang2025efficient} to achieve high performance.
In this paper, the fourth-order compact reconstruction using the same idea to reduce the computational cost is presented.

The structure of the paper is the following. Section 2 introduces the gas-kinetic scheme in an ALE formulation, including the governing equations on a moving mesh, the third-order gas-kinetic flux, and the determination of the grid velocity. Section 3 discusses the simplified fourth-order compact reconstruction and the GENO process for discontinuous. In Section 4, several test cases are used to validate the proposed method. Finally, the conclusion is presented in the last section.

\section{Compact Gas-Kinetic Scheme in the Arbitrary Lagrangian-Eulerian (ALE) Formulation}
\subsection{Governing equations on moving mesh}
In the finite volume method, the computational domain $\Omega$ is discretized into a series of non-overlap elements $\Omega_i$,
\begin{equation*}
\Omega =\bigcup \Omega_i,\ \Omega_i \bigcap \Omega_j=\phi (i \neq j).
\end{equation*}
The boundary of element $\Omega_i$ can be expressed as
\begin{equation*}
\partial \Omega_i=\bigcup_{p=1}^{N_f}\Gamma_{ip},
\end{equation*}
where $N_f$ is the number of element surface and $\Gamma_{ip}$ is surface of the element.
The Reynolds transport theorem gives the conservation laws in an arbitrary Lagrangian-Eulerian (ALE) formulation on a moving cell
\begin{equation*}
\frac{\td}{\td t}\int_{\Omega_i (t)}\mathbf{W} \td \Omega + \int_{\partial \Omega_i(t)} (\mathbf{F}(\mathbf{W})-\mathbf{W}\mathbf{U})\cdot \mathbf{n}\td S =0 ,
\end{equation*}
where $\mathbf{U}$ is the mesh velocity, $\mathbf{F}(\mathbf{W})$ is the flux passing through the fixed element surface, and $\mathbf{W}\mathbf{U}$ presents the flux caused by mesh motion.
For Navier-Stokes equations, $\mathbf{W}=[\rho,\rho V_1,\rho V_2,\rho V_3,\rho E]^T$ is the conservative flow variables, $\rho$ the density,
$\mathbf{V}=(V_1, V_2, V_3)$ the velocity, and $E$ the total energy of fluid.
\begin{itemize}
\item When $\mathbf{U}=\mathbf{V}$, the scheme is Lagrangian form;
\item When $\mathbf{U}=\mathbf{0}$, the scheme is Eulerian form.
\end{itemize}
A unified scheme can also be developed by choosing a mesh velocity between the above two limits \cite{hui2012computational}.

In this paper, only hexahedral elements are considered.  For simplicity, the grid velocity is assumed constant over a time interval. For three-dimensional computations, the control points may become non-coplanar even with a constant grid velocity, which introduces additional difficulties in preserving the order of accuracy and the geometric conservation law. Thus, a trilinear interpolation is introduced for a hexahedron cell with non-coplanar control points
\begin{equation*}
\mathbf{x}(\xi,\eta,\zeta)=\sum_{i=1}^{8} \mathbf{x}i \lambda_i(\xi,\eta,\zeta),
\end{equation*}
where $(\xi,\eta,\zeta)\in[-1,1]^3$ are the local coordinates, $\mathbf{x}i$ is the vertex of the hexahedron cell and $\lambda_i(\xi,\eta,\zeta)$ is the Lagrange basis function as follows:
\begin{equation*}
\lambda_i(\xi,\eta,\zeta)=\frac{1}{8}(1+\xi_i\xi)(1+\eta_i\eta)(1+\zeta_i\zeta),
\end{equation*}
where $(\xi_i,\eta_i,\zeta_i)$ are the $i$-th local coordinate of the vertex $\mathbf{x}i$.
With the trilinear interpolation, the triple integral over a hexahedron cell can be transformed into a single integral over the local coordinate as follows:
\begin{equation*}
\begin{aligned}
\iiint{\Omega} f(\mathbf{x}) \td\mathbf{x} &= \int_{-1}^{1} \int_ {-1}^{1} \int_{-1}^{1} f(\mathbf{x}(\xi,\eta,\zeta)) \left| \frac{\partial (x_1,x_2,x_3)}{\partial (\xi,\eta,\zeta)} \right| \td\xi \td\eta \td\zeta\\
& = \sum_{l,m,n=1}^{n} f(\mathbf{x}(\xi_l,\eta_m,\zeta_n)) \left| \frac{\partial (x_1,x_2,x_3)}{\partial (\xi,\eta,\zeta)} \right|\omega(\xi_l,\eta_m,\zeta_n) ,
\end{aligned}
\end{equation*}
where $\mathbf{x}(\xi,\eta,\zeta)$ is the position of the cell center in the local coordinate, $\omega(\xi_l,\eta_m,\zeta_n)$ is the Gaussian integration weight. For each cell interface, the trilinear interpolation reduces to a bilinear interpolation in the local coordinate, defined as follows:
\begin{equation*}
\mathbf{x}(\xi,\eta)=\sum_{i=1}^{4} \mathbf{x}_i \lambda_i(\xi,\eta),
\end{equation*}
where $(\xi,\eta)\in[-1,1]^2$ are the local coordinates, $\mathbf{x}_i$ is the vertex of the quadrilateral cell and $\lambda_i(\xi,\eta)$ is the Lagrange basis function as follows:
\begin{equation*}
\lambda_i(\xi,\eta)=\frac{1}{4}(1+\xi_i\xi)(1+\eta_i\eta),
\end{equation*}
where $(\xi_i,\eta_i)$ are the $i$-th local coordinate of the vertex $\mathbf{x}_i$.

Under this condition, the conservation law in the ALE formulation can be written as
\begin{equation*}
\mathbf{W}^{n+1}|\Omega^{n+1}|- \mathbf{W}^{n}|\Omega^{n}| =-\sum_{p=1}^{6}\int_{t^n}^{t^{n+1}} \int_{\Gamma_{p}}(\mathbf{F}(\mathbf{W})-\mathbf{W}\mathbf{U})\cdot \mathbf{n}\td S \td t,
\end{equation*}
where the time step is $\Delta t=t^{n+1}-t^n$ and $\Gamma_{p}$ is the $p$-th surface of the element. With the bilinear interpolation, the normal vector of the surface can be written as
\begin{equation*}
\mathbf{n}= \mathbf{x}_{,\xi}\times\mathbf{x}_{,\eta}=\frac{\partial \mathbf{x}}{\partial \xi} \times \frac{\partial \mathbf{x}}{\partial \eta}.
\end{equation*}
The surface integration can be written as
\begin{equation*}
\int_{\Gamma_{p}}(\mathbf{F}(\mathbf{W})-\mathbf{W}\mathbf{U})\cdot \mathbf{n}\td S=\int_0^1\int_0^{1}(\mathbf{F}(\mathbf{W})-\mathbf{W}\mathbf{U})\cdot (\mathbf{x}_{,\xi}\times\mathbf{x}_{,\eta}) \td \xi \td \eta.
\end{equation*}
Assuming the grid velocity is time-independent in one time interval, the coordinates of points on the surface can be written as $\mathbf{x}=\mathbf{x}^n+\mathbf{U}(t-t^n)$, where $\mathbf{x}^0$ is the initial position of the points.  If the surface value of the flow variables is also time-independent, the terms of flux caused by mesh motion can be calculated by
\begin{equation*}
\begin{aligned}
&\int_{t^n}^{t^{n+1}}\int_0^1\int_0^{1}\mathbf{W}\mathbf{U}\cdot (\mathbf{x}_{,\xi}\times\mathbf{x}_{,\eta}) \td \xi \td \eta \td t\\
=&\Delta t\int_0^1\int_0^{1}\mathbf{W}\mathbf{U} \cdot ((\mathbf{x}^n+\frac{1}{2}\mathbf{U}\Delta t)_{,\xi}\times(\mathbf{x}^n+\frac{1}{2}\mathbf{U}\Delta t)_{,\eta}) \td \xi \td \eta\\
+&\frac{1}{12}\Delta t^3\int_0^1\int_0^{1}\mathbf{W}\mathbf{U} \cdot (\mathbf{U}_{,\xi}\times\mathbf{U}_{,\eta}) \td \xi \td \eta, \
\end{aligned}
\end{equation*}
which means that the flux crossing the surface can be calculated by two parts: the flux calculated by the midpoint of the mesh motion and a geometric correction term. Thus, the total flux can be approximated as
\begin{equation*}
\begin{aligned}
\int_{t^n}^{t^{n+1}}\int_{\Gamma_{p}}(\mathbf{F}(\mathbf{W})-\mathbf{W}\mathbf{U})\cdot \mathbf{n}\td S \td t
\approx&\int_{t^n}^{t^{n+1}}\int_0^1\int_0^{1}(\mathbf{F}(\mathbf{W})-\mathbf{W}\mathbf{U}) \cdot \mathbf{n}_{1/2} \td \xi \td \eta \td t\\
+&\frac{1}{12}\Delta t^3\int_0^1\int_0^{1}\mathbf{W}^*\mathbf{U} \cdot (\mathbf{U}_{,\xi}\times\mathbf{U}_{,\eta}) \td \xi \td \eta ,
\end{aligned}
\end{equation*}
where $\mathbf{n}_{1/2}=(\mathbf{x}^n+1/2\mathbf{U}\Delta t)_{,\xi}\times(\mathbf{x}^n+1/2\mathbf{U}\Delta t)_{,\eta}$ and $\mathbf{W}^{*}$ is the flow variables in the surface at the beginning of the time interval, which will be obtained at following instructions. Thus, the final form of the finite volume method for ALE formulation can be written as
\begin{equation}\label{fvm_ale}
\begin{aligned}
\mathbf{W}^{n+1}|\Omega^{n+1}|- \mathbf{W}^{n}|\Omega^{n}| = -\sum_{p=1}^{6} &\left[\int_{t^n}^{t^{n+1}} \int_0^1\int_0^{1}(\mathbf{F}(\mathbf{W})-\mathbf{W}\mathbf{U})\cdot \mathbf{n}_{1/2} \td \xi \td \eta \td t \right.\\
&\left.+\frac{1}{12}\Delta t^3\int_0^1\int_0^{1}\mathbf{W}^*\mathbf{U} \cdot (\mathbf{U}_{,\xi}\times\mathbf{U}_{,\eta}) \td \xi \td \eta \right],
\end{aligned}
\end{equation}
where the surface integral is approximated by the Gaussian integration with $2\times 2$ points at each surface.
\subsection{BGK equation in ALE Formulation}
The gas-kinetic BGK equation can be written as
\begin{equation*}
\frac{\partial f}{ \partial t}+ \mathbf{v}\cdot \nabla_x f  = \frac{g-f}{\tau},
\end{equation*}
where $f=f(\mathbf{x},t,\mathbf{v},\xi)$ is the gas distribution function, $g$ is the corresponding equilibrium state, and $\tau$ is the collision time.
$\mathbf{v}=(v_1,v_2,v_3)$ is the particle velocity.
The collision term in the above equation describes the evolution process from a non-equilibrium state to an equilibrium one with
the satisfaction of the compatibility condition
\begin{equation*}
\int \frac{g-f}{\tau} \Psi \td\mathbf{v}\td\Xi = 0,
\end{equation*}
where $\Psi=(1,v_1,v_2,v_3,\frac{1}{2}(v_1^2+v_2^2+v_3^2+\xi^2))^T$ and $\td\Xi=\td\xi_1\cdots\td\xi_K$ ($K$ is the number of internal degree of freedom, i.e. $K=(5-3\gamma)/(\gamma-1)$ for three-dimensional flow and $\gamma$ is the specific heat ratio). According to the Chapman-Enskog Expansion for the BGK equation, the macroscopic governing equations can be derived \cite{vki1998}. In the continuum region, the BGK equation can be rearranged, and the gas distribution function can be expanded as
\begin{equation*}
f=g-\tau \mathcal{D}{\mathbf{u}}g+\tau \mathcal{D}{\mathbf{u}}(\tau\mathcal{D}{\mathbf{u}})g-\tau \mathcal{D}{\mathbf{u}}[ \tau \mathcal{D}{\mathbf{u}}(\tau\mathcal{D}{\mathbf{u}})]g+\cdots,
\end{equation*}
where $\mathcal{D}{\mathbf{u}}=\frac{\partial }{\partial t}+ \mathbf{u}\cdot \nabla$. With the zeroth-order truncation $f=g$, the Euler equations can be obtained. For the first-order truncation $f=g-\tau \mathcal{D}{\mathbf{u}}g$, the Navier-Stokes equations with the dynamic  viscosity coefficient $\mu=\tau p$ ($p$ is pressure) and Prandtl number $\text{Pr}=1$ can be obtained.

In this paper, a high-order gas-kinetic scheme will be constructed in the arbitrary Lagrangian-Eulerian (ALE) framework for three-dimensional flow. The particle speed related to grid $\mathbf{w}=(w_1,w_2,w_3)$ is
\begin{equation*}
\mathbf{w} = \mathbf{v}-\mathbf{U},
\end{equation*}
where $\mathbf{U}=(U_1,U_2,U_3)$ is the grid velocity. The BGK equation in a moving reference frame can be written as
\begin{equation} \label{BGK}
\frac{\partial f}{ \partial t}+ \mathbf{w}\cdot \nabla_x f  = \frac{g-f}{\tau}.
\end{equation}

The flux of ALE-formulation finite volume method Eq. (\ref{fvm_ale}) can be evaluated by
\begin{equation*}
\mathbb{F}(\mathbf{W},\mathbf{x}^{1/2} ) =(\mathbf{F} (\mathbf{W})-\mathbf{W}\mathbf{U}) \cdot \mathbf{n}_{1/2}=\int f(\mathbf{x},t,\mathbf{w},\xi)\mathbf{w}\cdot \mathbf{n}_{1/2} \mathbf{ \Psi} \td\mathbf{v}\td\Xi.
\end{equation*}
In the actual computation, the numerical flux can be obtained by taking moments of the gas distribution function with the related velocity
\begin{equation}\label{flux}
\mathbb{F}^\prime(\mathbf{W},\mathbf{x}^{1/2} ) =\int f(\mathbf{x},t,\mathbf{w},\xi)\mathbf{w}\cdot \mathbf{n}_{1/2} \mathbf{ \Psi}^\prime \td\mathbf{v}\td\Xi,
\end{equation}
where ${\mathbf{ \Psi}^\prime} =(1,{w}_1,{w}_2,{w}_3,\frac{1}{2}({w}_1^2+{w}_2^2+{w}_3^2+\xi^2))^T$. The component of $\mathbb{F}$ can be determined by the combination of $\mathbb{F}^\prime$ as follows
\begin{equation*}
\begin{cases}
\mathbb{F}_\rho=\mathbb{F}_\rho^\prime ,\\
\mathbb{F}_{\rho V_1}=\mathbb{F}_{\rho W_1}^\prime + U_1 \mathbb{F}_\rho^\prime ,\\
\mathbb{F}_{\rho V_2}=\mathbb{F}_{\rho W_2}^\prime + U_2 \mathbb{F}_\rho^\prime ,\\
\mathbb{F}_{\rho V_3}=\mathbb{F}_{\rho W_3}^\prime + U_3 \mathbb{F}_\rho^\prime ,\\
\mathbb{F}_{\rho E}=\mathbb{F}_{\rho E}^\prime+U_1\mathbb{F}_{\rho W_1}^\prime +U_2\mathbb{F}_{\rho W_2}^\prime +U_3\mathbb{F}_{\rho W_3}^\prime + \frac{1}{2}(U_1^2+U_2^2+U_3^2) \mathbb{F}_\rho^\prime . \\
\end{cases}
\end{equation*}
By defining the rotating matrix $\mathbf{T}=\text{diag}(1,\mathbf{T}^\prime,1)$ and
\begin{equation*}
\mathbf{T}^\prime =\left( \begin{matrix}
n_1&n_2&n_3\\
-n_2&n_1+\frac{n_3^2}{1+n_1}&-\frac{n_2n_3}{1+n_1}\\
-n3&-\frac{n_2n_3}{1+n_1}&1-\frac{n_3^2}{1+n_1}
\end{matrix}\right),n_1\neq -1,
\end{equation*}
and when $n_1=-1$, $\mathbf{T}^\prime$ becomes $\text{diag}(-1,-1,1)$, the flux (Eq. \ref{flux}) can be evaluated by
\begin{equation*}
\mathbb{F}^\prime(\mathbf{W},\mathbf{x}^{1/2} ) =\mathbf{T}^{-1}\int f(\tilde{\mathbf{x}}_{p,k},t,\tilde{\mathbf{w}},\xi) \tilde{w}_1 \tilde{\mathbf{ \Psi}}^\prime \td\tilde{\mathbf{w}}\td\Xi,
\end{equation*}
where the origin point of local coordinate is $\tilde{\mathbf{x}}=(0,0,0)$  with $x$-direction in $\mathbf{n}$ and $\widetilde{\mathbf{ \Psi}}^\prime =(1,\tilde{w}_1,\tilde{w}_2,\tilde{w}_3,\frac{1}{2}(\tilde{w}_1^2+\tilde{w}_2^2+\tilde{w}_3^2+\xi^2))^T$. The microscopic velocities in local coordinate are given by $\tilde{\mathbf{w}} = \mathbf{T}^\prime \mathbf{w}$ and $\tilde{w}_1= n_1w_1 +n_2w_2+n_3w_3$.

\subsection{Gas evolution model}
In order to construct the numerical fluxes at $\mathbf{x} = (0,0,0)^T$, the integral solution of the BGK equation Eq.(\ref{BGK}) is
\begin{equation}\label{solbgk}
f(\mathbf{x},t, \mathbf{w},\xi)=\frac{1}{\tau}\int_0^t g(\mathbf{x}^{\prime},t^{\prime},\mathbf{w},\xi)e^{-(t-t^{\prime})/\tau}\td t^{\prime}+ e^{t/\tau}f_0(\mathbf{x}_{0},\mathbf{w}),
\end{equation}
with the trajectory
\begin{equation*}
\mathbf{x}=\mathbf{x}^{\prime}+\mathbf{w}(t-t^{\prime}).
\end{equation*}
In Eq.(\ref{solbgk}), $f_0$ is the initial gas distribution function, and $g$ is the corresponding equilibrium state. $\mathbf{x}_0$ are the initial position by tracing back particles $\mathbf{x}$ at time $t$ back to $t=0$.

With the consideration of possible discontinuity at an interface, the initial distribution is constructed as
\begin{equation}\label{ini_gas}
f_0(\mathbf{x}_{0},\mathbf{w}_0)=f_0^l(\mathbf{x}_{0},\mathbf{w}) (1 -\mathbb{H}(x_1)) +f_0^r(\mathbf{x}_{0},\mathbf{w})\mathbb{H}(x_1),
\end{equation}
where $\mathbb{H}$ is the Heaviside function. $f_0^{l}$ and $f_0^{r}$ are the initial gas distribution functions on the left and right sides of the interface, which are determined by corresponding initial macroscopic variables and their spatial derivatives.
With the second-order accuracy, $f_0^k(\mathbf{x},\mathbf{w})$ is constructed by Taylor expansion around $(\mathbf{x},\mathbf{w})$
\begin{equation}\label{taylor_ex}
f_0^k(\mathbf{x})=f_G^k (\mathbf{0}) +\frac{\partial f_G^k}{\partial x_i}x_i+\frac{1}{2}\frac{\partial^2 f_G^k}{\partial x_i\partial x_j}x_ix_j
\end{equation}
for $k=l,r$. Based on the Chapman-Enskog expansion, $f_G^k$ is given by
\begin{equation}\label{ce_ex}
f_G^k= g^k[1-\tau(\frac{\partial g^k}{\partial t}+  \frac{\partial g^k}{\partial x_i}w_i)],
\end{equation}
where $g^k$ is the equilibrium distribution function defined by the macroscopic variables $\mathbf{W}^k$ at both sides of a cell interface.
Then, the equilibrium distribution is defined by the Taylor expansion
\begin{equation}\label{equdis}
\begin{aligned}
g=g^c+\frac{\partial g^c}{\partial x_i}x_i+\frac{\partial g^c}{\partial t}t+\frac{1}{2}\frac{\partial^2 g^c}{\partial x_i\partial x_j}x_ix_j+\frac{\partial^2 g^c}{\partial x_i\partial t} x_i t+\frac{1}{2}\frac{\partial^2 g^c}{\partial t^2}t^2
\end{aligned}
\end{equation}

In the smooth flow region, the collision time for viscous flow is determined by $\tau=\mu/p$, where $\mu$ is the dynamic viscosity coefficient and $p$ is the pressure at the cell interface, and for inviscid flow, the collision time should be set as zero. To properly capture the unresolved shock structure, additional numerical dissipation is required. The physical collision time $\tau$ in the exponential function part can be replaced by a numerical collision time $\tau_n$. For inviscid flow, it is set as
\begin{equation*}
\tau_n=C_1\Delta t + C_2\frac{|p_l-p_r|}{p_l+p_r}\Delta t ,
\end{equation*}
and for viscous flow, it is
\begin{equation*}
\tau_n=\tau + C_2\frac{|p_l-p_r|}{p_l+p_r}\Delta t ,
\end{equation*}
where $p_l$ and $p_r$ denote the pressure on the left and right sides of the cell interface. In this paper, we have $C_1=0.01, C_2=5.0$.

By substituting  Eq.(\ref{taylor_ex}), Eq.(\ref{ce_ex}) and Eq.(\ref{equdis}) into Eq.(\ref{solbgk}) with $\tau$ and $\tau_n$, the third-order time-dependent gas distribution function can be obtained. However, the full third-order time-dependent gas distribution function is too complex to use in numerical implementation. A simplified version has been proposed by Zhou et al. \cite{zhou2017simplification}. The new set of coefficients is introduced as
\begin{equation*}
g a_{x_i}=\frac{\partial g}{\partial x_i},\quad g a_{t}=\frac{\partial g}{\partial t}, \quad g a_{x_i x_j}=\frac{\partial^2 g}{\partial x_i\partial x_j}, \quad g a_{x_i t}=\frac{\partial^2 g}{\partial x_i\partial t}, \quad g a_{tt}=\frac{\partial^2 g}{\partial t^2}.
\end{equation*}
These coefficients can be obtained by
\begin{equation*}
\begin{aligned}
&<a_{x_i}>=\frac{\partial \mathbf{W}}{\partial x_i},\quad <a_t+a_{x_i}w_i>=0,\\
&<a_{x_i x_j}>=\frac{\partial^2 \mathbf{W}}{\partial x_i\partial x_j}, \quad <a_{x_i t}+a_{x_ix_j}w_j>=0, \quad <a_{tt}+a_{x_i t}w_i>=0,
\end{aligned}
\end{equation*}
where $<\cdots>=\int (\cdots)g \Psi\td\mathbf{w}\td\Xi$ denotes the moment of the distribution function. And the equilibrium distribution function and its spatial derivatives are defined as
\begin{equation*}
\begin{aligned}
\int g^c\Psi\td\mathbf{w}\td\Xi&=\mathbf{W}^c=\int_{w_1>0}g^l\Psi\td\mathbf{w}\td\Xi+\int_{w_1<0}g^r\Psi\td\mathbf{w}\td\Xi,\\
\int a_{x_i} g^c \td\mathbf{w}\td\Xi&=\int_{w_1>0}a_{x_i}g^l \td\mathbf{w}\td\Xi+\int_{w_1<0}a_{x_i}g^r\td\mathbf{w}\td\Xi,\\
\int a_{x_i x_j} g^c \td\mathbf{w}\td\Xi&=\int_{w_1>0}a_{x_i x_j}g^l \td\mathbf{w}\td\Xi+\int_{w_1<0}a_{x_i x_j}g^r\td\mathbf{w}\td\Xi.
\end{aligned}
\end{equation*}
The final simplified gas distribution function is given by
\begin{equation}\label{solution}
\begin{aligned}
f\left(\mathbf{0}, t, \mathbf{w}, \xi\right)
&=g^c\left[1+a_t t+\frac{1}{2}a_{tt}t^2-\tau(a_t+a_{x_i}w_i+a_{tt}t+a_{x_i t}w_it)\right] \\
&-e^{-t/\tau_n}g^c(1-a_{x_i}w_it)\\
&+e^{-t/\tau_n}g^l(1-a_{x_i}w_it)H(w_1) \\
&+e^{-t/\tau_n}g^r(1-a_{x_i}w_it)(1-H(w_1))
\end{aligned}
\end{equation}
The fluxes in Eq.(\ref{flux}) can be obtained by taking the moments of the above distribution function. The calculation of moments is described in \cite{xuGKS2001}. And the surface conservative value $\mathbf{W}^{*}$ in Eq.(\ref{fvm_ale}) is obtain by taking moments of equilibrium state $g^c$.

\subsection{Evolution of the cell-averaged spatial gradients}
By taking moments of the above gas distribution function in Eq. (\ref{solution}), the time-accurate conservative flow variables in the moving frame at the Gaussian points can also be obtained
\begin{equation*}
\mathbf{W}_{p_t,k}^\prime (t^{n+1})= \mathbf{T}^\prime\left(\int  \widetilde{\mathbf{ \Psi}} f(\tilde{\mathbf{x}}_{p_t,k},t^{n+1},\tilde{\mathbf{w}},\xi) \td\mathbf{w}\td\Xi \right),
\end{equation*}
by which the corresponding conservative flow variables in the absolute inertia frame $\mathbf{W}_{p_t,k} (t^{n+1})$ can be obtained.
According to the Divergence theorem, the cell-averaged gradients over cell $\Omega_i$ at time $t^{n+1}$ are
\begin{equation*}
\overline{\nabla \mathbf{W}}^{n+1}_i = \frac{2}{|\Omega_i|^n+ |\Omega_i^{n+1}|}\sum_{p=1}^{6}\int{\Gamma_{p}} \mathbf{W}^{n+1} \mathbf{n}_p dS,
\end{equation*}
where the surface integration can be calculated by Gaussian quadrature
\begin{equation*}
\int{\Gamma_{p}} \mathbf{W}^{n+1}  \mathbf{n}_{p} dS = \int_0^1\int_0^{1} \mathbf{W}^{n+1} \mathbf{n}_{1/2} \td \xi \td \eta \approx \sum_{k=1}^{4}|S_{p}| \omega_k \mathbf{W}_{p,k}(t^{n+1})\mathbf{n}_{p,k}.
\end{equation*}
Thus, both the conservative flow variables and the cell-averaged gradients can be obtained by the above procedure. The compact reconstruction can be constructed in the next section.
\subsection{Mesh Motion Formulation}
This article considers two types of mesh motion when the mesh velocity is not given directly. The first method is the variational approach, which is based on the gradient of the flow variables, and it is used to adapt the mesh to the flow. The second method is the Lagrangian velocity method, which is based on the cell-centered Lagrangian nodal solver. It can track the moving shock waves and is used to handle the strong shock waves. In addition, to avoid the mesh distortion, a smoothing process is applied after the mesh velocity is determined.
\subsubsection{Variational Approach}
The mesh velocity can be determined by the variational approach \cite{tang2003adaptive}, and the corresponding Euler-Lagrange equations can be obtained from
\begin{equation*}
\begin{aligned}
(\omega x_{,\xi})_{,\xi} + (\omega x_{,\eta})_{,\eta} + (\omega x_{,\zeta})_{,\zeta} = 0, \\
(\omega y_{,\xi})_{,\xi} + (\omega y_{,\eta})_{,\eta} + (\omega y_{,\zeta})_{,\zeta} = 0, \\
(\omega z_{,\xi})_{,\xi} + (\omega z_{,\eta})_{,\eta} + (\omega z_{,\zeta})_{,\zeta} = 0,
\end{aligned}
\end{equation*}
where $(x, y,z)$ and $(\xi,\eta,\zeta)$ denote the physical and computational coordinates, $\omega$ is the monitor function and can be chosen as a function of the flow variables, such as the density, velocity, pressure, or their gradients. In numerical tests, without a special statement, the monitor function takes the form
\begin{equation*}
\omega = \max\{2,\sqrt{1 + \alpha |\nabla \rho|^2}\}.
\end{equation*}
To conveniently parallelize the computation, the finite element scheme is used to discretize the equations, and the PESTc library\cite{petsc-efficient} is used to solve the resulting linear system. The mesh distribution can be directly generated, and the grid velocity is obtained by
\begin{equation*}
\mathbf{U}_{ijk} = \frac{x_{n+1} - x_n}{\Delta t}.
\end{equation*}
More details can be found in \cite{tang2003adaptive}.

\subsubsection{Lagrangian Velocity Method}
The mesh velocity can be determined by the cell-centered Lagrangian nodal solver \cite{Lagrangianvel}. The basic idea is to solve two half-one-dimensional Riemann problems at each cell interface by assuming that the velocity at each interface equals the point velocity. We denote $C(p)$ as the set of cells $c$ that share the common vertex $p$ and $F_p(c)$ as the set of faces of cell $c$ that share the common vertex $p$. The grid speed $\mathbf{U}_p$ can be given by
\begin{equation*}
\mathbf{U}_p=\mathbb{M}_p^{-1}\sum_{c\in C(p)}\sum_{f\in F_p(c)}[S_fp_c\mathbf{N}_f^c+\mathbb{M}_{pcf}\mathbf{V}_c],
\end{equation*}
where $p_c$ and $\mathbf{V}+c$ are the pressure and velocity of cell $c$ and $S_f, \mathbf{N_f}$ is the surface area and normal direction. The matrix $\mathbb{M}_{pcf}$ is calculated by
\begin{equation*}
\mathbb{M}_{pcf}=S_f\rho_c a_c(\mathbf{N}_f^c \otimes \mathbf{N}_f^c),
\end{equation*}
where $a_c$ and $\rho_c$ are the sound speed and density of cell $c$ and $\otimes$ is the operator of tensor production, and matrix $\mathbb{M}_p$ is the sum of $\mathbb{M}_{pcf}$ for faces and is a symmetric positive definite matrix. In this case, the coordinates of points in the next time step can be calculated by
\begin{equation*}
\mathbf{x}_p^{n+1}=\mathbf{x}_p^n+\mathbf{U}_p\Delta t.
\end{equation*}
With the Lagrangian velocity, the meshes become distorted, and computation may break down. Therefore, the meshes need to be smoothed by
\begin{equation*}
\widetilde{\mathbf{x}}_p=(1-\omega)\mathbf{x}_p+\omega \frac{1}{N_p}\sum_{ip\in N(p)}\mathbf{x}_{ip},
\end{equation*}
where $N(p)$ is the set of vertexes jointing with vertex $p$, $N_p$ is the size of $N(p)$, and $\omega$ is relaxation coefficient.

\subsection{The Algorithm of ALE Method}
Algorithm \ref{alg1} presents the complete computational algorithm, with bold text indicating the special treatments compared to the Eulerian formulation. The mesh velocity is determined once per time step. This necessitates updating the cell volume, the normal vector, the area, and the coordinates of the Gaussian points at interfaces at every step. Using the single-stage high-order time-integration method, the geometry variables are computed only once per step.
\begin{algorithm}
%\textsl{}\setstretch{1.8}
\renewcommand{\algorithmicrequire}{\textbf{Input:}}
\renewcommand{\algorithmicensure}{\textbf{Output:}}
\caption{CGKS in ALE formulation}
\label{alg1}
\begin{algorithmic}[1]
\WHILE{the computation uncompleted}
\STATE calculate the time step $\Delta t$ according to CFL number
\STATE {\bf{ determine mesh velocity}}
\STATE define boundary condition for ghost cell
\STATE reconstruct the cell distribution
\STATE evolution for interfaces
\STATE {\bf{update the geometry }}
\STATE update cell average conservative values and the first-order spatial derivatives of conservative values
\ENDWHILE
\end{algorithmic}
\end{algorithm}

\section{Fourth-order generalized ENO reconstruction with compact stencils}
In this section, the generalized ENO (GENO) reconstruction\cite{zhao2025generalized,yang2025effective} is presented. The GENO method employs an adaptive mechanism that transitions from high-order linear reconstruction in smooth flow regions to second-order reconstruction near discontinuities. This method strikes an optimal balance between high accuracy--maintained by high-order linear reconstruction across a broad range of wavenumbers--and shock-capturing robustness, which is guaranteed by preserving the ENO property.
\subsection{Fourth-order linear compact reconstruction for large stencil}

\begin{figure}[hbt!]
\centering
 \subfigure[Large stencil]{ \label{largeStencil} \includegraphics[width=0.3\textwidth]{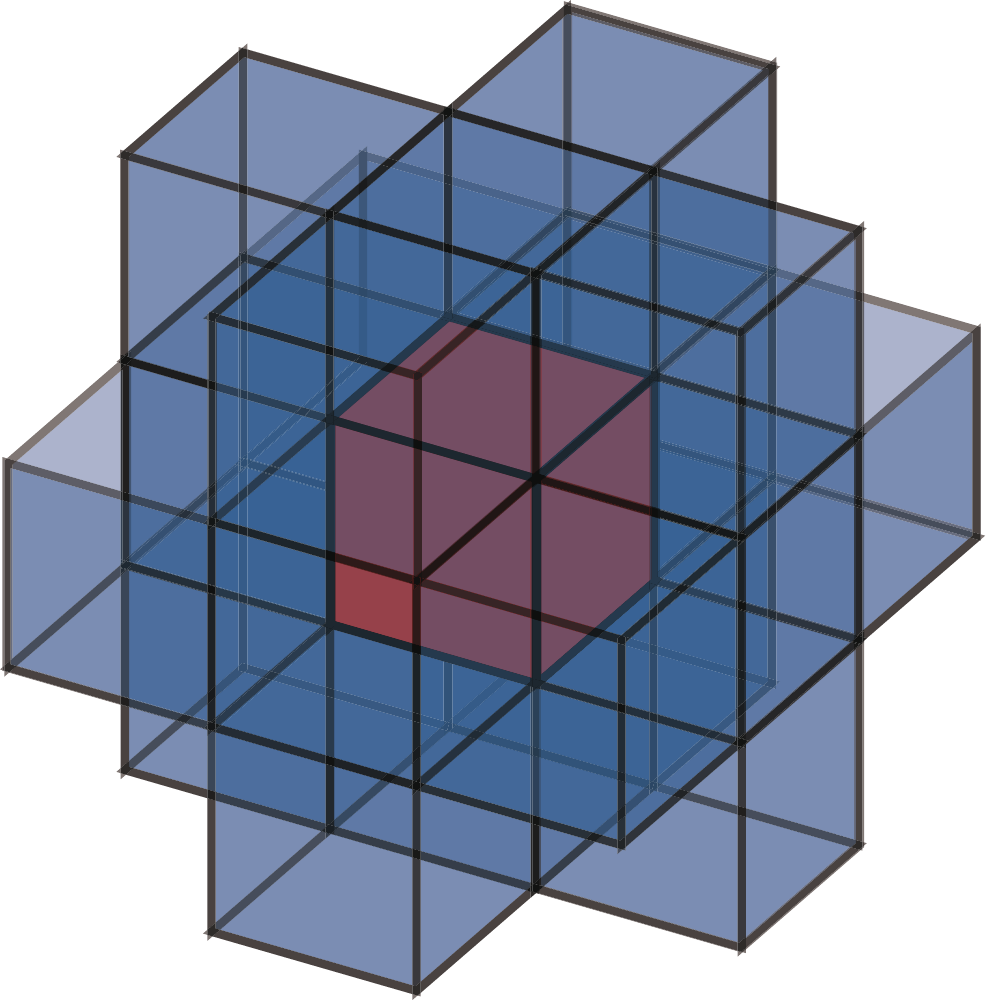}}
\quad \subfigure[Eight second-order sub-stencils]{ \label{subStencil} \includegraphics[width=0.6\textwidth]{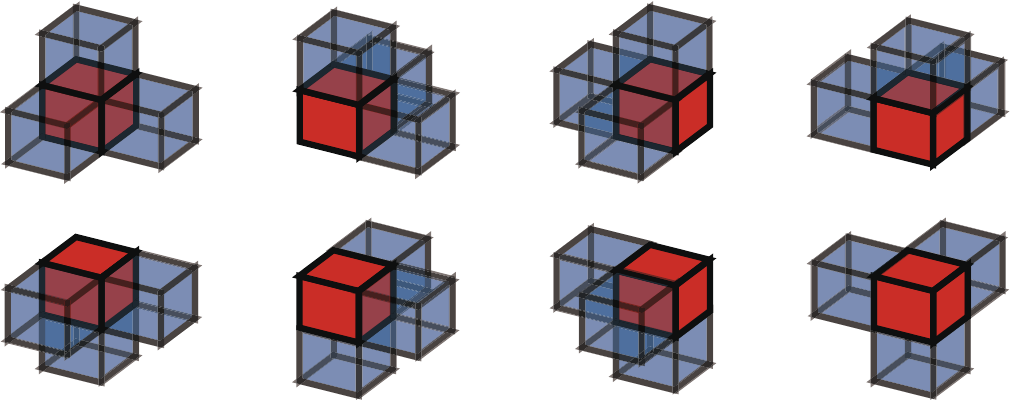}}
 \caption[]{The reconstruction stencils of fourth-order reconstruction}
\end{figure}

For the target cell $\Omega_i$, the large stencils of the compact reconstruction are shown in Figure \ref{largeStencil}, which includes the target cell, the face-neighbor cells, and the edge-neighbor cells. Based on this compact reconstruction stencil, a fifth-order compact GKS was first proposed in \cite{yang2025effective}. In that work, fifth-order compact reconstruction was achieved by projecting the cell-averaged gradients of edge-neighboring cells onto specific directions. The construction of the fourth-order linear compact reconstruction in the current study follows the approach developed in \cite{yang2025effective}. The only difference is that the target cell's line-averaged data is not required here to achieve fourth-order accuracy. Therefore, the details of the fourth-order compact reconstruction are not given here and can be referred to in \cite{yang2025effective}. In addition, to improve the robustness of the reconstruction, eight second-order polynomials $p_j^1$ ($j=1,\cdots,8$) are constructed by the sub-stencils shown in Figure \ref{subStencil}.

To maintain strong stability under conditions of extreme deformation and high irregularity of structured grids, this study employs reconstruction directly on the physical grid rather than on the computational grid \cite{yang2025effective}. 
For the stationary mesh, the reconstruction matrix can be calculated initially and stored in memory for subsequent use. For the moving mesh, the reconstruction matrix needs to be calculated in every time step. 
Since constructing the reconstruction matrix dominates the computational cost, conventional fourth-order reconstruction—which requires a full matrix—presents a performance bottleneck.
To improve the computational efficiency, reducing the size of the reconstruction matrix is a key issue. A simplified fourth-order compact reconstruction with small-size matrix approach is presented. 
The fourth-order linear reconstruction polynomial is given by
\begin{equation*}
 p^3(\tilde{\mathbf{x}})= \sum_{n_1+n_2+n_3=0}^{3} \frac{q_{n_1,n_2,n_3} }{n_1!n_2!n_3!} \tilde{x}_1 ^{n_1} \tilde{x}_2 ^{n_2} \tilde{x}_3^{n_3},
\end{equation*}
where $q_{n_1,n_2,n_3}$ are reconstruction coefficients, $\tilde{ \mathbf{x}}=(\tilde{x}_1,\tilde{x}_2,\tilde{x}_3)=(\mathbf{x}-\mathbf{x}_0)/h$ are the normalized coordinates of the target cell ($\mathbf{x}_0$ is the coordinate of cell center) and $h=|\Omega_0|/(\max_j S_j)$, ($|\Omega_0|$ is the volume of cell and $S_j$ is the area of cell's surface ) is the cell size. Because the gas-kinetic scheme can give the first-order spatial derivatives of the cell-averaged values, the first-order spatial derivatives can be reconstructed first,
\begin{equation*}
\begin{aligned}
 p^3_{,x_1}&= \frac{1}{2}q_{3,0,0}\tilde{x}_1^2+\frac{1}{2}q_{1,2,0}\tilde{x}_2^2+\frac{1}{2}q_{1,0,2}\tilde{x}_3^2+q_{2,1,0}\tilde{x}_1\tilde{x}_2+q_{2,0,1}\tilde{x}_1\tilde{x}_3+q_{1,1,1}\tilde{x}_2\tilde{x}_3+O(\tilde{x}_i),\\
 p^3_{,x_2}&=\frac{1}{2}q_{2,1,0}\tilde{x}_1^2+ \frac{1}{2}q_{0,3,0}\tilde{x}_2^2+\frac{1}{2}q_{0,1,2}\tilde{x}_3^2+q_{1,2,0}\tilde{x}_1\tilde{x}_2+q_{1,1,1}\tilde{x}_1\tilde{x}_3+q_{0,2,1}\tilde{x}_2\tilde{x}_3+O(\tilde{x}_i),\\
 p^3_{,x_3}&= \frac{1}{2}q_{2,0,1}\tilde{x}_1^2+\frac{1}{2}q_{0,2,1}\tilde{x}_2^2+\frac{1}{2}q_{0,0,3}\tilde{x}_3^2+q_{1,1,1}\tilde{x}_1\tilde{x}_2+q_{1,0,2}\tilde{x}_1\tilde{x}_3+q_{0,1,2}\tilde{x}_2\tilde{x}_3+O(\tilde{x}_i),\\
 \end{aligned}
\end{equation*}
where $O(\tilde{x}_i)$ denotes the zero-order and first-order terms. Meanwhile, the distribution function of cell-averaged slopes can be written by,
\begin{equation*}
 p^3_{,x_i}= \frac{1}{2}b_{i,0}\tilde{x}_1^2+\frac{1}{2}b_{i,1}\tilde{x}_2^2+\frac{1}{2}b_{i,2}\tilde{x}_3^2+b_{i,3}\tilde{x}_1\tilde{x}_2+b_{i,4}\tilde{x}_1\tilde{x}_3+b_{i,5}\tilde{x}_2\tilde{x}_3+O(\tilde{x}_i),
\end{equation*}
thus, we can obtain the third-order reconstruction coefficients $q_{i,j,k}$ by 
\begin{equation*}
\begin{aligned}
 &q_{3,0,0}=b_{1,0},\quad q_{0,3,0}=b_{2,0},\quad q_{0,0,3}=b_{3,0},\\
 &q_{2,1,0}=(b_{1,3}+b_{2,0})/2,\quad q_{1,2,0}=(b_{1,1}+b_{2,3})/2,\quad q_{1,0,2}=(b_{1,2}+b_{3,4})/2,\\
 &q_{2,0,1}=(b_{1,4}+b_{3,0})/2,\quad q_{0,2,1}=(b_{1,5}+b_{3,1})/2,\quad q_{0,1,2}=(b_{2,2}+b_{3,5})/2,\\
 &q_{1,1,1}=(b_{1,5}+b_{2,4}+b_{3,3})/3.
\end{aligned}
\end{equation*}
After obtaining the third-order reconstruction coefficients, the lower-order coefficients can be determined by 
\begin{equation*}
\iiint_{\Omega_i}\sum_{n_1+n_2+n_3=0}^{2}\frac{q_{n_1,n_2,n_3}}{n_1!n_2!n_3!} \tilde{x}_1 ^{n_1} \tilde{x}_2 ^{n_2} \tilde{x}_3^{n_3}\td V = |\Omega_i|Q_i- \iiint_{\Omega_i}\sum_{n_1+n_2+n_3=3}\frac{q_{n_1,n_2,n_3}}{n_1!n_2!n_3!} \tilde{x}_1 ^{n_1} \tilde{x}_2 ^{n_2} \tilde{x}_3^{n_3}\td V.
\end{equation*}
The reconstruction strictly preserves the target cell's averaged value while employing least-squares minimization for neighboring cells. Notably, both the complete polynomial and gradient distributions share identical reconstruction matrices. Compared with the full reconstruction matrix \cite{yang2025effective} ($20\times61$), only small matrix with dimension $10\times19$ was used in above reconstruction.

\subsection{Nonlinear GENO weights}
 The generalized ENO (GENO) reconstruction can be written as
\begin{equation*}
 R(\tilde{\mathbf{x}})=\chi p^3(\tilde{\mathbf{x}})-(1-\chi)\sum_{j=1}^8\bar{\omega}_jp_j^1(\tilde{\mathbf{x}}),
\end{equation*}
which can be seen as a convex combination of the fourth-order polynomial and the second-order ENO reconstruction based on the sub-stencils, through a path function $\chi$ of the GENO reconstruction scheme\cite{zhao2025generalized,yang2025effective}. The ENO weights are calculated by
 \begin{equation*}
 \omega_{j}=\left(\frac{1}{\epsilon+\text{IS}_j}\right)^5,\bar{\omega}_{j}=\frac{\omega_{j}}{\sum \omega_{j}},j=0,\cdots,8
\end{equation*}
where $\epsilon=10^{-16}$ is a small positive constant and $\text{IS}_j$ is the smoothness indicator of the $j$-th sub-stencil. The smoothness indicator is calculated by
\begin{equation*}
\text{IS}_j=\sum_{|\alpha|=1}^{3}\Omega^{\frac{2}{3}|\alpha|-1} \iiint_{\Omega}\left(D^{\alpha} p_j^1(\mathbf{x})\right)^{2} \td V,
\end{equation*}
where $\alpha$ is a multi-index, $D^{\alpha}=\partial^{|\alpha|}/\partial x_1^{\alpha_1}\partial x_2^{\alpha_2}\partial x_3^{\alpha_3}$ is the derivative operator and $|\alpha|=\alpha_1+\alpha_2+\alpha_3$. The path function $\chi$ is calculated by
 \begin{equation*}
 \chi = \tanh (C\alpha)/\tanh(C),C=20,
\end{equation*}
 where
\begin{equation*}
\alpha = \frac{2\alpha^H}{\alpha^H+\alpha^L},\alpha^{H,L}=1+\left(\frac{\text{IS}^\tau}{\text{IS}^{H,L}+\epsilon}\right)^5.
 \end{equation*}
And the smoothness indicators are calculated by the fourth-order polynomial via
\begin{equation*}
\begin{aligned}
\text{IS}^H=\sum_{|\alpha|=1}^{3}\Omega^{\frac{2}{3}|\alpha|-1} \iiint_{\Omega}\left(D^{\alpha} p^{3}(\mathbf{x})\right)^{2} \td V\approx\sum_{|\alpha|=1}^{3} \Omega^{\frac{2}{3}|\alpha|} q_{\alpha_1,\alpha_2,\alpha_3}^2/h^{2|\alpha|}, \\
\text{IS}^L=\sum_{|\alpha|=1}\Omega^{\frac{2}{3}|\alpha|-1} \iiint_{\Omega}\left(D^{\alpha} p^{3}(\mathbf{x})\right)^{2} \td V\approx\sum_{|\alpha|=1} \Omega^{\frac{2}{3}|\alpha|} q_{\alpha_1,\alpha_2,\alpha_3}^2/h^{2|\alpha|}, \\
\text{IS}^\tau=\sum_{|\alpha|=3}\Omega^{\frac{2}{3}|\alpha|-1} \iiint_{\Omega}\left(D^{\alpha} p^{3}(\mathbf{x})\right)^{2} \td V\approx\sum_{|\alpha|=3} \Omega^{\frac{2}{3}|\alpha|} q_{\alpha_1,\alpha_2,\alpha_3}^2/h^{2|\alpha|}. \\
\end{aligned}
\end{equation*}
The desired non-equilibrium states at Gaussian points become
\begin{equation*}
 Q_{p, k}^{l, r}=R^{l, r}\left(\mathbf{x}_{p, k}\right), \left(Q_{x_{i}}^{l, r}\right)_{p, k}=\frac{\partial R^{l, r}}{\partial x_{i}}\left(\mathbf{x}_{p, k}\right),\left(Q_{x_{i}x_j}^{l, r}\right)_{p, k}=\frac{\partial^2 R^{l, r}}{\partial x_{i}\partial x_j}\left(\mathbf{x}_{p, k}\right).
\end{equation*}

\section{ Numerical Experiments }
Both two-dimensional and three-dimensional cases are tested in this section. For two-dimensional problems, the three-dimensional solver is used with a single layer and a periodic boundary condition in the $z$ direction. The time step is given by $\Delta t= \min \Delta t_i$, where $\Delta t_i$ is defined in each cell
\begin{equation*}
\Delta t_i =C_{CFL}\min(\frac{h}{|V|_i+c_i},\frac{h^2}{3\nu_i}),
\end{equation*}
where $C_{\text{CFL}}$ is the CFL number, $|V|_i, c_i$, and $\nu_i=(\mu/\rho)_i$ is the magnitude of related velocities, sound speed, and kinematic viscosity coefficient of cell $i$. Here, we set the CFL number to 0.3 unless specified \cite{jiThreedimensionalCompactHighorder2020}.
\subsection{Accuracy Test}
The advection of density perturbation for three-dimensional flow is presented to test the order of accuracy. The computation domain is $[0,1]\times[0,1]\times[0,1]$. Four uniform meshes with $8^3, 16^3, 32^3, 64^3$ tetrahedral cells are used in this case. The flow at time $t$ is
\begin{equation*}
\begin{aligned}
\rho(x,y,z) = 1+0.2\sin(2\pi x_r)\sin(2\pi y_r)\sin(2\pi z_r) ,\\
p(x,y,z) =1,V_1=V_2=V_3=1,
\end{aligned}
\end{equation*}
where the $(x_r,y_r,z_r)$ are the reference coordinates which depend on time
\begin{equation*}
(x_r,y_r,z_r)=(x,y,z)-(1,1,1)t.
\end{equation*}
In order to validate the order of accuracy with moving meshes, the time-dependent moving meshes are considered in this form
\begin{equation*} \text{Motion 1:}\begin{cases}
x&=x_0+0.1\sin(2\pi t)\sin(2\pi x_0)\sin(2\pi y_0)\sin(2\pi z_0), \\
y&=y_0+0.1\sin(2\pi t)\sin(2\pi x_0)\sin(2\pi y_0)\sin(2\pi z_0), \\
z&=z_0+0.1\sin(2\pi t)\sin(2\pi x_0)\sin(2\pi y_0)\sin(2\pi z_0), \\
\end{cases}
\end{equation*}
\begin{equation*}
\text{Motion 2:}\begin{cases}
x&=x_0+0.05\sin(2\pi t)(\sin(2\pi x_0)+\sin(2\pi x_0)\sin(2\pi y_0)\sin(2\pi z_0)), \\
y&=y_0+0.05\sin(2\pi t)(\sin(2\pi y_0)+\sin(2\pi x_0)\sin(2\pi y_0)\sin(2\pi z_0)), \\
z&=z_0+0.05\sin(2\pi t)(\sin(2\pi z_0)+\sin(2\pi x_0)\sin(2\pi y_0)\sin(2\pi z_0)). \\
\end{cases}
\end{equation*}
The errors and numerical orders are shown in Table \ref{er_r}, indicating that fourth-order accuracy is achieved under the moving mesh.
\begin{table}[htb!]
\centering
\caption{The errors and accuracy in moving mesh}\label{er_r}
\begin{tabular}{ccccc c c}
\hline
mesh    &  $Error{L^1}$& $O{L^1}$  &  $Error_{L^2}$   &$O_{L^2}$        & $Error_{L^\infty}$&$O_{L^\infty}$  \\ \hline
Motion 1 \\
$8^3$   &7.07E-03 &    & 8.98E-03&     & 2.22E-02&        \\
$16^3$  &3.55E-04 &4.31& 4.65E-04& 4.27& 1.72E-03 &3.69 \\
$32^3$  &2.27E-05 &3.97& 3.27E-05& 3.83& 3.52E-04 &2.29 \\
$64^3$  &1.60E-06 &3.83& 3.08E-06& 3.41& 5.86E-05 &2.59 \\
\hline
Motion 2 \\
$8^3$ &9.01E-03&      &1.15E-02&      &3.98E-02&     \\ 
$16^3$  &5.82E-04& 3.95 &7.64E-04& 3.92 &4.01E-03& 3.31 \\
$32^3$  &3.72E-05& 3.97 &4.82E-05& 3.99 &2.93E-04& 3.78 \\
$64^3$  &2.71E-06& 3.78 &3.72E-06& 3.69 &5.82E-05& 2.33 \\
\hline
\end{tabular}
\end{table}

As a reference, the accuracy test results for the stationary mesh are shown in Table \ref{ersta}, which indicate that the errors of the moving and fixed meshes are comparable.
\begin{table}[htb!]
\centering
\caption{The errors and accuracy in stationary mesh}\label{ersta}
\begin{tabular}{ccccccc}
\hline
mesh    &  $Error_{L^1}$& $O_{L^1}$  &  $Error_{L^2}$   &$O_{L^2}$       & $Error_{L^\infty}$&$O_{L^\infty}$  \\ \hline
$8^3$   &7.07E-03&&  8.98E-03 && 2.22E-02  \\
$16^3$   &3.55E-04& 4.31 &4.65E-04& 4.27 &1.72E-03& 3.69 \\
$32^3$   &2.27E-05& 3.97 &3.27E-05& 3.83 &3.52E-04& 2.29 \\
$64^3$   &1.60E-06& 3.83 &3.08E-06& 3.41 &5.86E-05& 2.59 \\
\hline
\end{tabular}
\end{table}

Next, the geometric conservation law, which governs the maintenance of uniform flow through the moving mesh, is tested. The same mesh and the same mesh-moving method are used in the test. The initial condition is set as
\begin{equation*}
\rho(x,y,z) = 1,
p(x,y,z) =1,V_1=V_2=V_3=1.
\end{equation*}
The density errors at $t=1$ are shown in Table \ref{errgcl}. The results indicate that the errors are reduced to machine zero, thereby satisfying the geometric conservation law.
\begin{table}[htb!]
\centering
\caption{The errors of geometric conservation law in moving mesh}\label{errgcl}
\begin{tabular}{cccc}
\hline
mesh    &  $Error_{L^1}$&  $Error_{L^2}$        & $Error_{L^\infty}$  \\ \hline
Motion 1 \\
$8^3$   &1.03E-15  &1.28E-15 &4.22E-15\\
$16^3$   &1.34E-15  &1.68E-15 &5.66E-15\\
$32^3$   &1.79E-15  &2.24E-15 &8.99E-15\\
$64^3$   &2.68E-15  &3.39E-15 &1.82E-14\\
\hline
Motion 2 \\
$8^3$   &1.01E-15 &1.27E-15 &4.22E-15\\
$16^3$  &1.33E-15 &1.67E-15 &6.44E-15\\
$32^3$  &1.82E-15 &2.28E-15 &8.99E-15\\
$64^3$  &2.74E-15 &3.48E-15 &1.82E-14\\
\hline
\end{tabular}
\end{table}

Finally, the computational efficiency of the current reconstruction was compared with that of the previous fourth-order reconstruction \cite {yang2025effective}. The testing platform consists of an AMD Ryzen Threadripper 2990WX CPU operating at 4.2 GHz, featuring 32 cores and 128 GB of memory. The code is compiled with GCC 11.4.0 using the \texttt{-O2} optimization flag and linked with Open MPI 4.1.5. 32 cores used for the test case. The mesh motion, defined as motion 1, was used for this test case. The computational time is shown in Table \ref{computationTime}, which shows that the computational efficiency of the current reconstruction is about 2.4x to 3.0x faster than the previous reconstruction.
\begin{table}[htb!]
\centering
\caption{Efficiency comparison: the total execution time and speedup for current and previous reconstruction for advection of density perturbation }\label{computationTime}
\begin{tabular}{cccc}
\hline
mesh    &   current reconstruction(s)& Previous reconstruction(s)& Speedup \\ \hline
$16^3$   &20  &49 &2.45\\
$32^3$   &292  &789 &2.70\\
$64^3$   &4236 &12472 &2.94\\
\hline
\end{tabular}
\end{table}

\subsection{Double shear layer}

The double shear layer is a canonical test problem for a scheme's accuracy and resolution in incompressible flows. A comparison was conducted by Minion et al. \cite{minion1997performance} for the solution at different resolutions. The computational domain is $[0,1]\times[0,1]$, and the periodic boundary condition is used in both $x$ and $y$ directions. Total mesh elements are $192\times192$. The initial flow is set as 
\begin{equation*}
  \begin{aligned}
 & V_1(x,y)= \begin{cases}
    \tanh{(k(y-0.25))},&y\leq 0.5, \\
    \tanh{(k(0.75-y))},&y>0.5,
  \end{cases} \\
 & V_2(x,y)=\delta \sin{(2\pi x)},
  \rho(x,y)=1,p(x,y)=\rho/(\gamma M^2), M=0.15, \gamma=1.4,
  \end{aligned}
\end{equation*}
where $k=100$ and $\delta=0.05$. The kinetic viscosity is $\nu=5.0\times10^{-5}$. Linear reconstruction is used for this case. The variational approach is used for mesh deformation, and the monitor function for mesh deformation is set as
\begin{equation*}
  \omega=\sqrt{1+0.01|\partial_y V_1- \partial_x V_2 |^2}.
\end{equation*}

\begin{figure}[htbp!]
  \centering
  \subfigure[Vorticity contour on ALE mesh]{\label{fig:doubleShearlineContorALE}\includegraphics[width=0.4\textwidth]{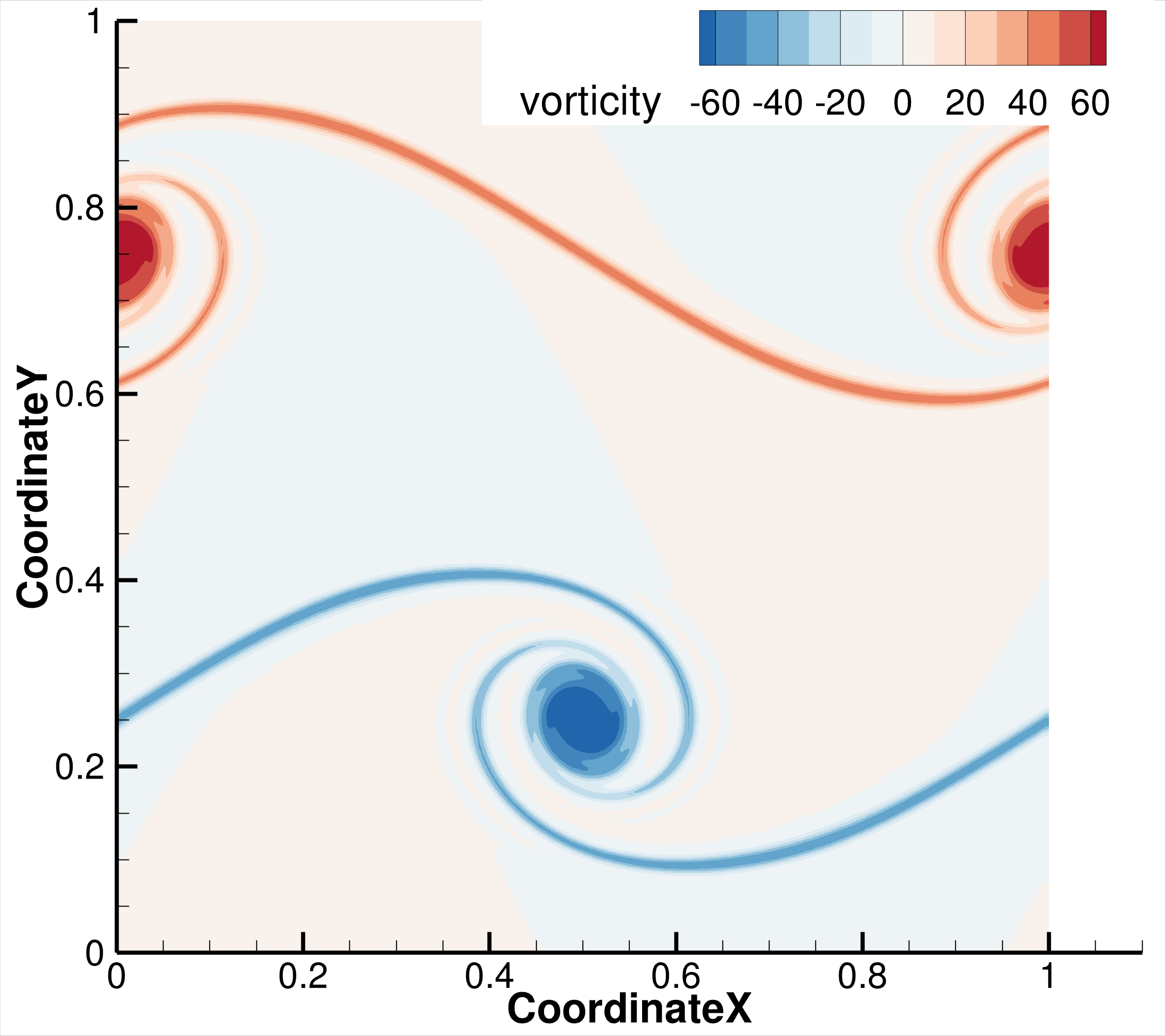}}\quad
  \subfigure[Mesh deformation]{\label{fig:doubleShearlineMeshALE}\includegraphics[width=0.4\textwidth]{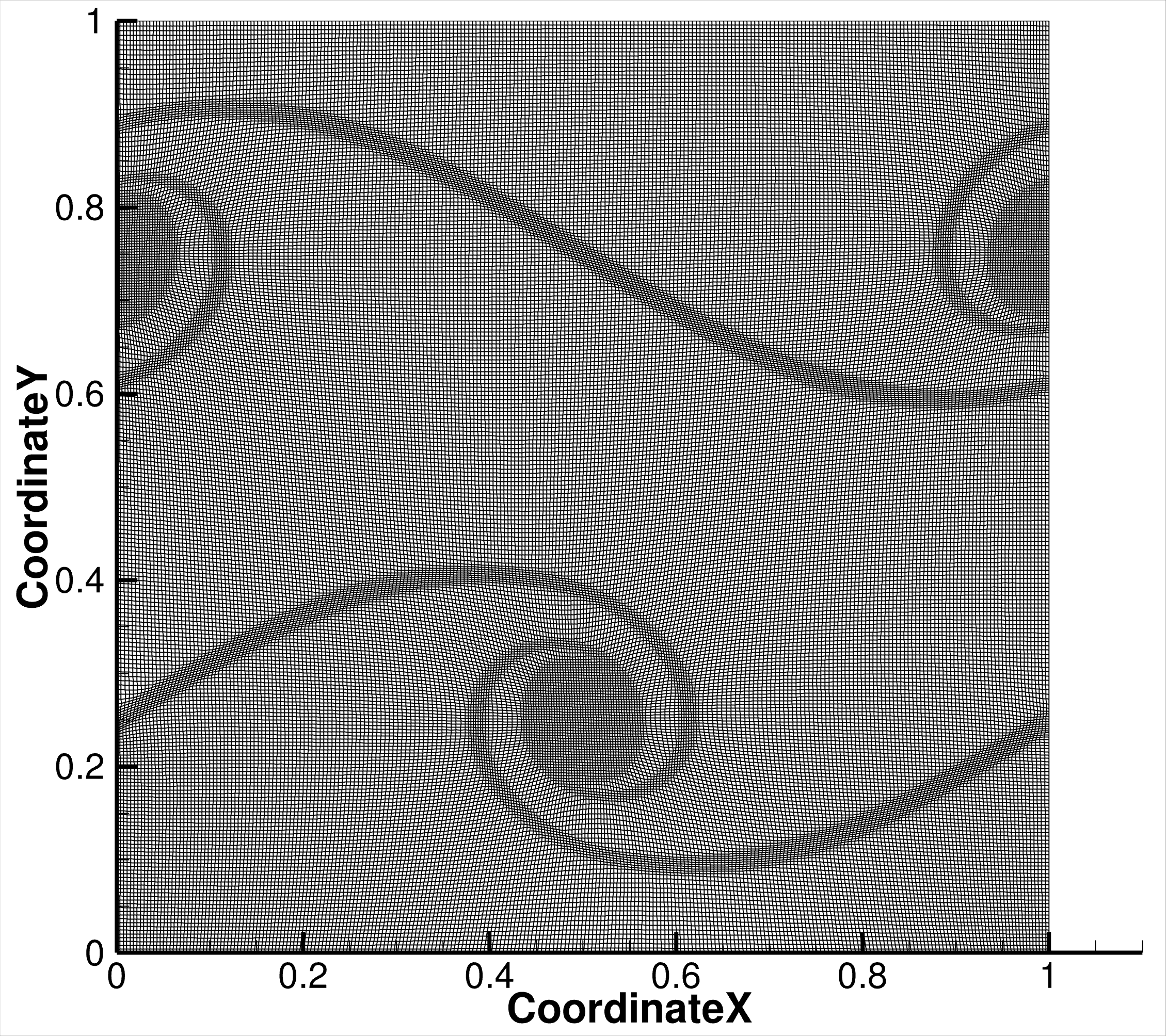}}  
  \caption{The solution of double shear layer with $192^2$ mesh elements at time $t=0.8$}\label{fig:doubleShearline}
\end{figure}

\begin{figure}[htbp!]
  \centering
  \includegraphics[width=0.4\textwidth]{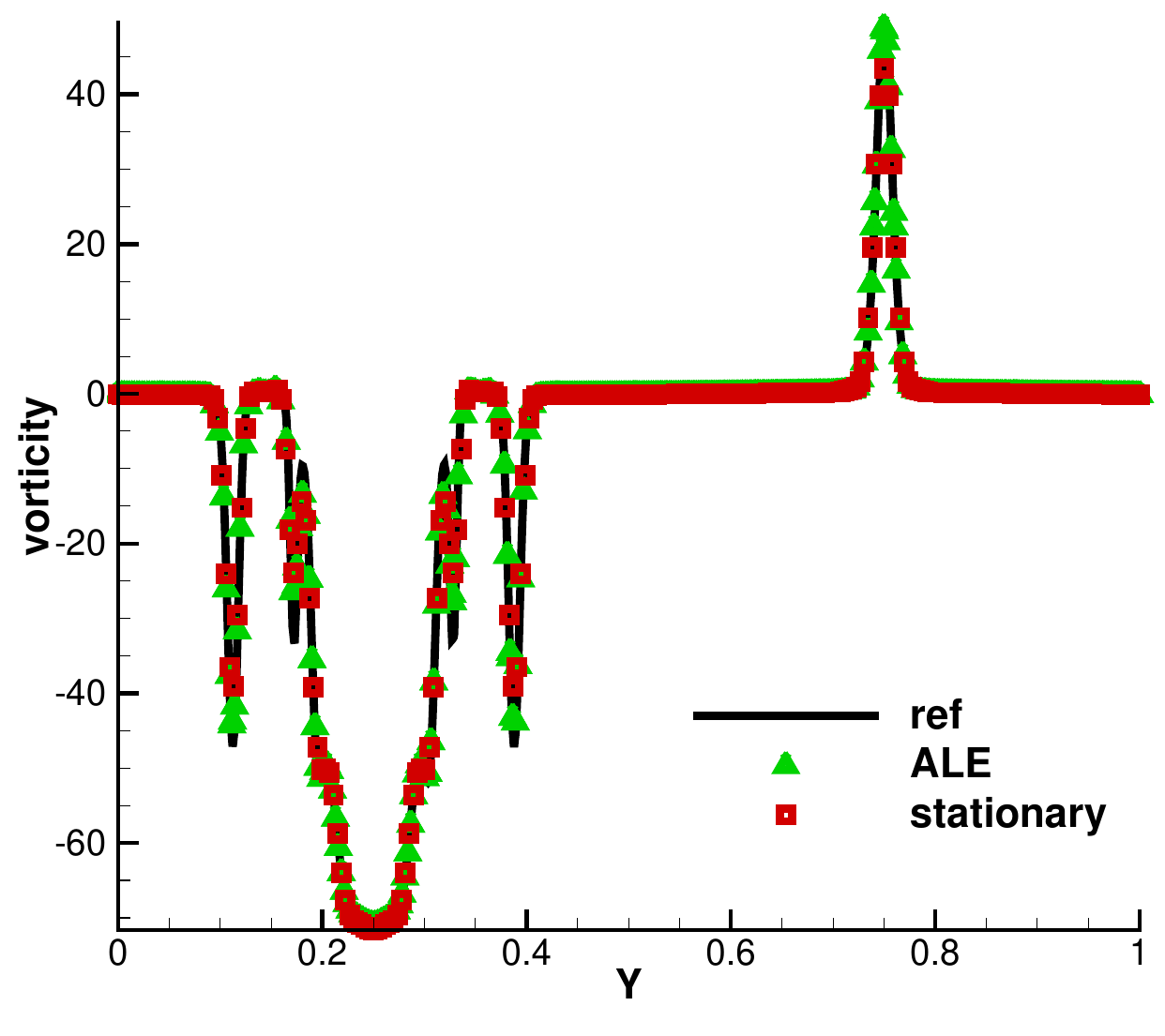}\quad
  \includegraphics[width=0.4\textwidth]{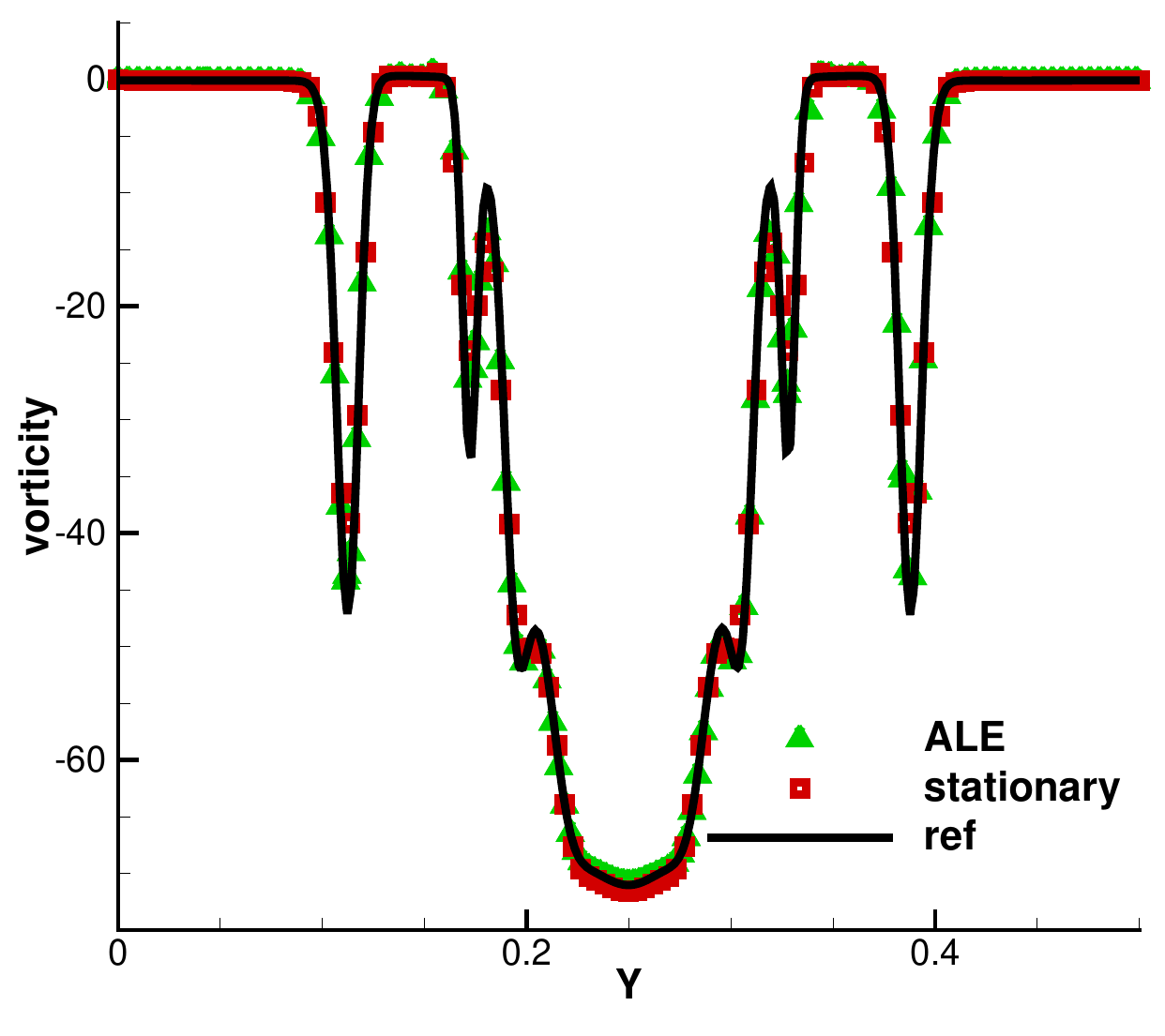}
  \caption{The vorticity distribution along the line $x=0.5$}\label{fig:doubleShearlineVorticity}
\end{figure}

The solution of time $t=0.8$ is shown in Figure \ref{fig:doubleShearline}. The vorticity contours on ALE mesh are shown in Figure \ref{fig:doubleShearlineContorALE}. The mesh deformation is shown in Figure \ref{fig:doubleShearlineMeshALE}, which shows that the mesh is adapted and refined in the shear layer region.
The vorticity distribution along the line $x=0.5$ is shown in Figure \ref{fig:doubleShearlineVorticity}, which shows that the vorticity is concentrated in the shear layer region. A comparison between ALE and stationary-mesh results shows that the ALE method better resolves the shear layer than the stationary-mesh method, particularly at the vorticity extrema, due to mesh adaptation via the monitor function.

\subsection{Two-dimensional Riemann problem}
In this subsection, two standard two-dimensional Riemann problems for the Euler equations are simulated. The computational domain is $[0,1] \times[0,1]$. The mesh resolution is $400 \times400$. A variational mesh deformation approach is employed, with deformation coefficient $\alpha=0.01$.

\begin{figure}[htbp!]
  \centering
  \subfigure[Stationary mesh]{\includegraphics[width=0.4\textwidth]{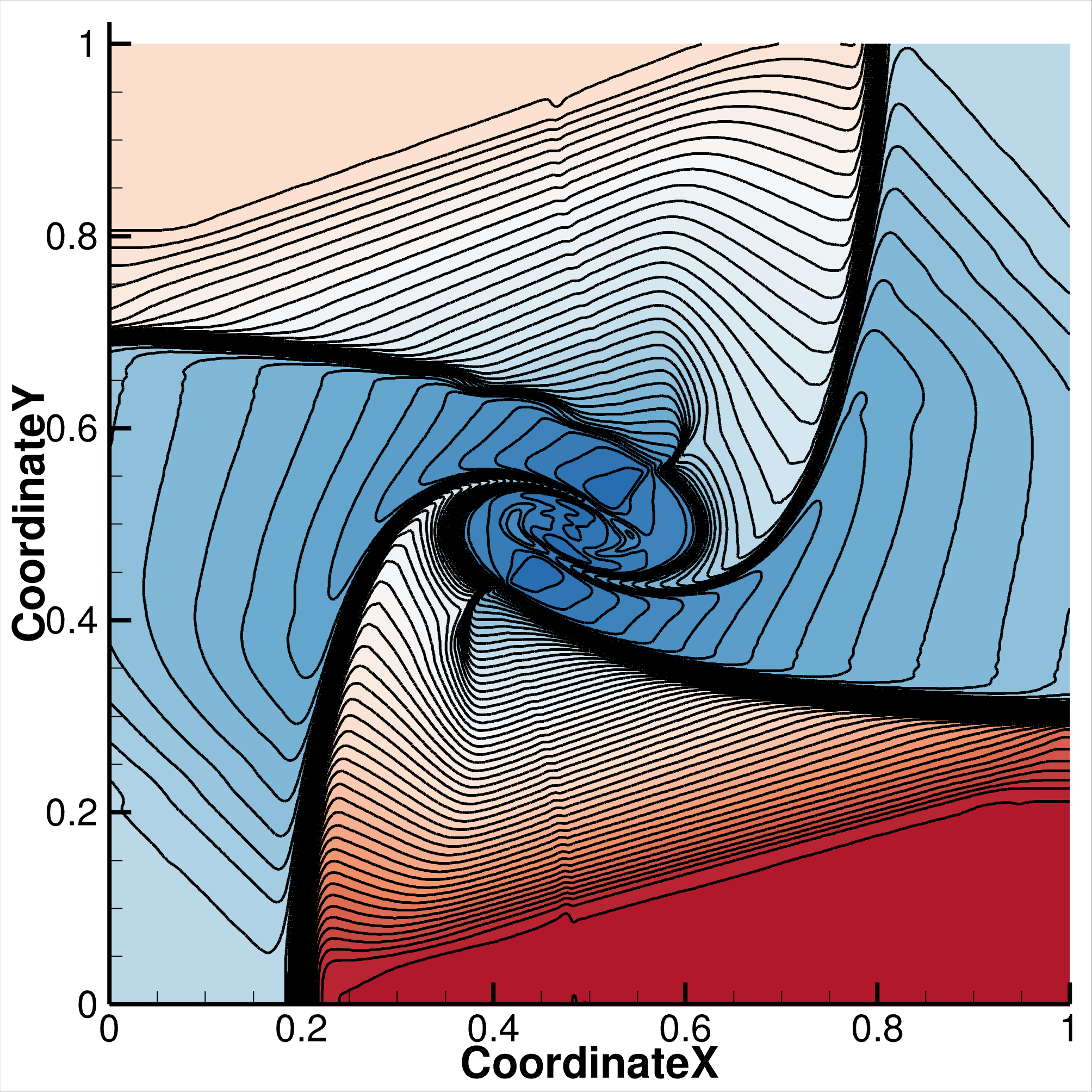}}\quad
  \subfigure[ALE mesh]{\includegraphics[width=0.4\textwidth]{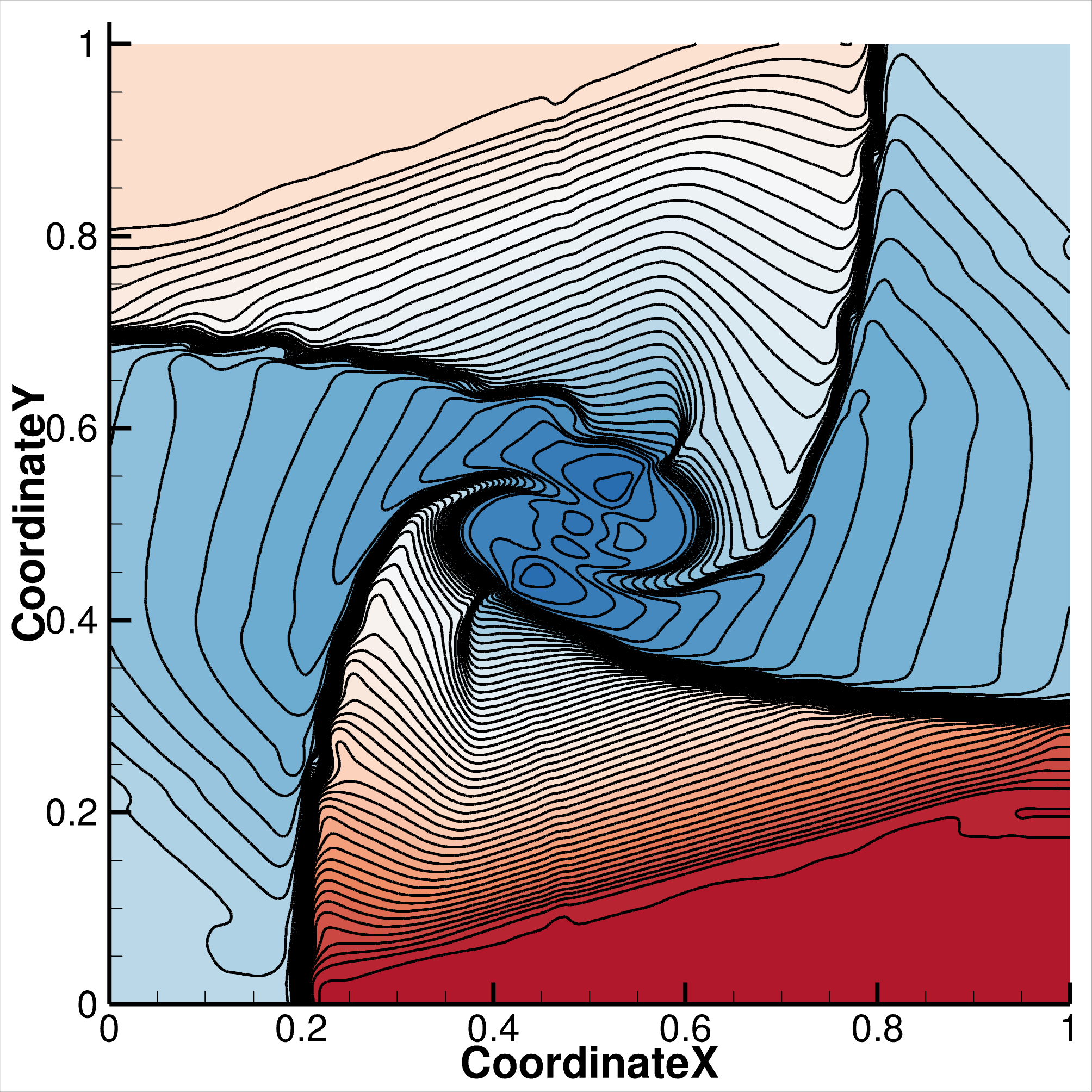}}
  \caption{The density contours (contour levels from 0.2 to 3.0 with increments of 0.05) of a two-dimensional Riemann problem with contact discontinuities at time $t=0.4$}\label{fig:riemann2Contor}
\end{figure}

\begin{figure}[htbp!]
  \centering
  \subfigure[Mesh deformation]{\includegraphics[width=0.4\textwidth]{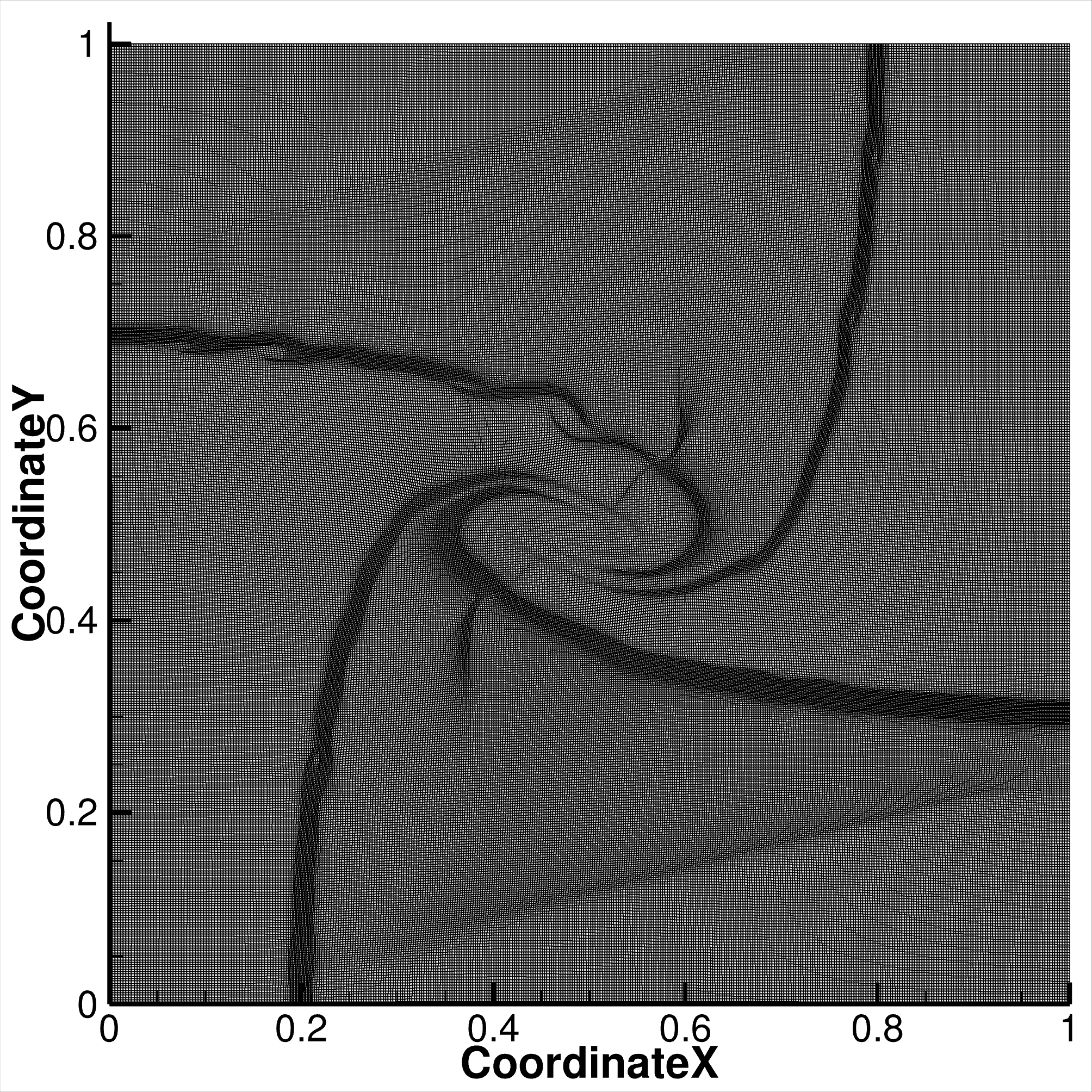}}\quad
  \subfigure[Mesh deformation enlargement]{\includegraphics[width=0.4\textwidth]{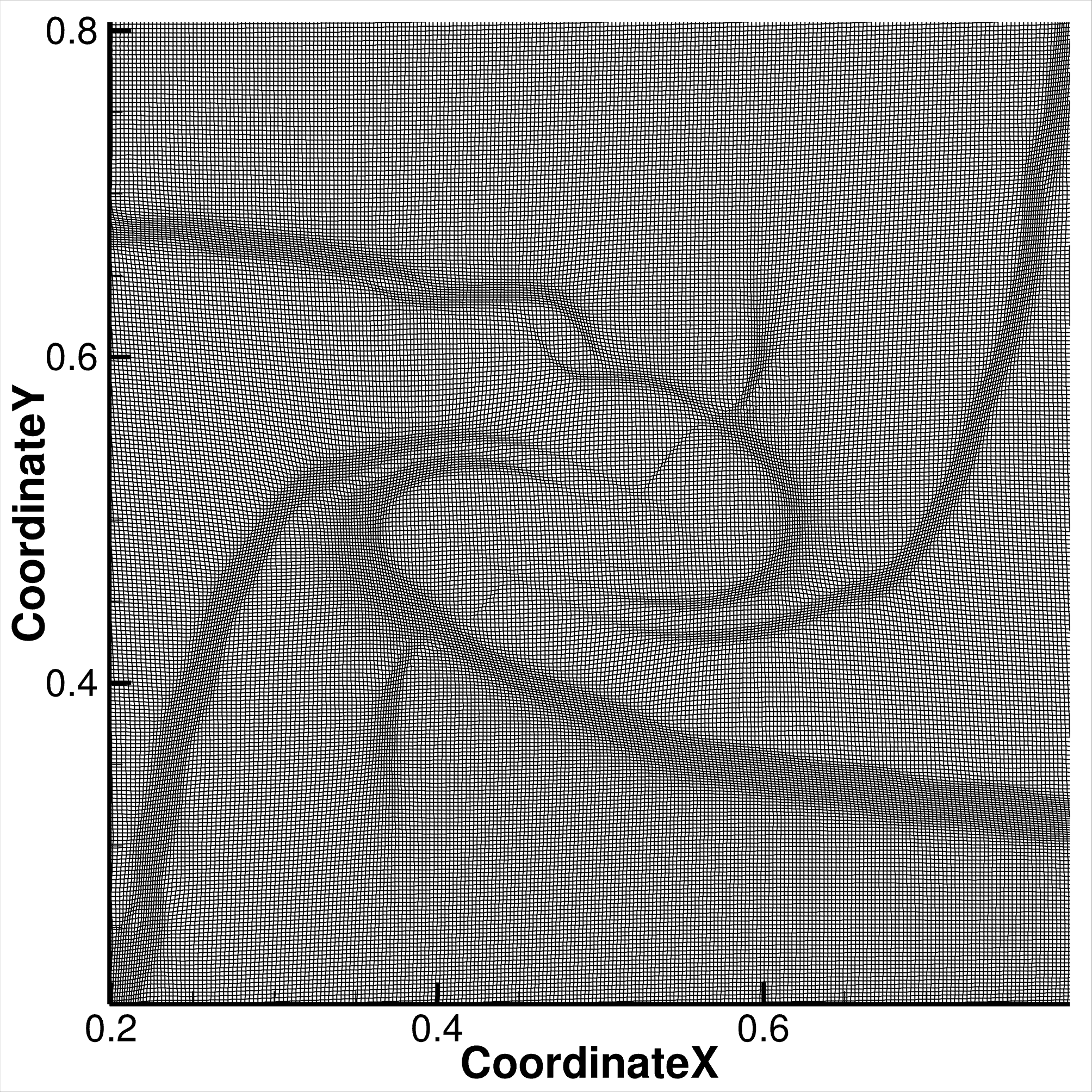}}
  \caption{The mesh distribution of two-dimensional Riemann problem with contact discontinuity at time $t=0.4$}\label{fig:riemann2Mesh}
\end{figure}

The first test case is the Riemann problem featuring a contact discontinuity, which serves as a standard benchmark for the Euler equations. The initial conditions are specified as
\begin{equation*}
  (\rho,V_1,V_2,p)=\begin{cases}
   (1,0.75,-0.5,1),        &x>0.5,y>0.5, \\
   (2,0.75,0.5,1), &x<0.5,y>0.5, \\
   (1,-0.75,0.5,1),      &x<0.5,y<0.5, \\
   (3,-0.75,-0.5,1), &x>0.5,y<0.5. \\
  \end{cases}
\end{equation*}
This case simulates shear instabilities among four initial contact discontinuities. The solution at time $t=0.4$ is shown in Figure \ref{fig:riemann2Contor}, which shows the shear instabilities among four initial contact discontinuities. A comparison between ALE and stationary-mesh results shows that the ALE solution is unstable at the contact discontinuity, indicating higher resolution than the stationary mesh.
The mesh distribution on the ALE grid and a local magnification at time $t=0.4$ are shown in Figure \ref{fig:riemann2Mesh}, illustrating that the mesh is adaptively refined in the shear instability zone.

\begin{figure}[htbp!]
  \centering
  \subfigure[Stationary mesh]{\includegraphics[width=0.4\textwidth]{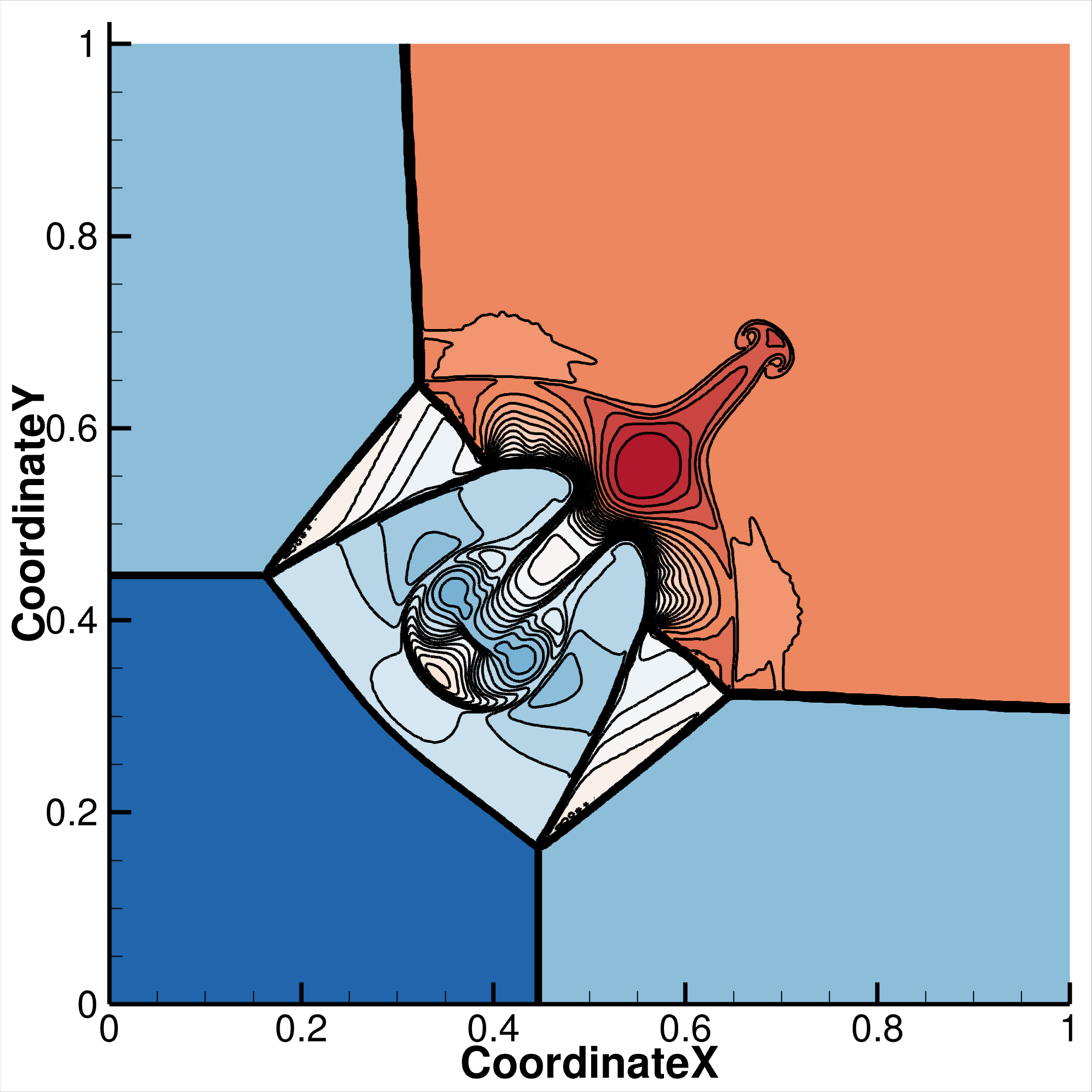}}\quad
  \subfigure[ALE mesh]{\includegraphics[width=0.4\textwidth]{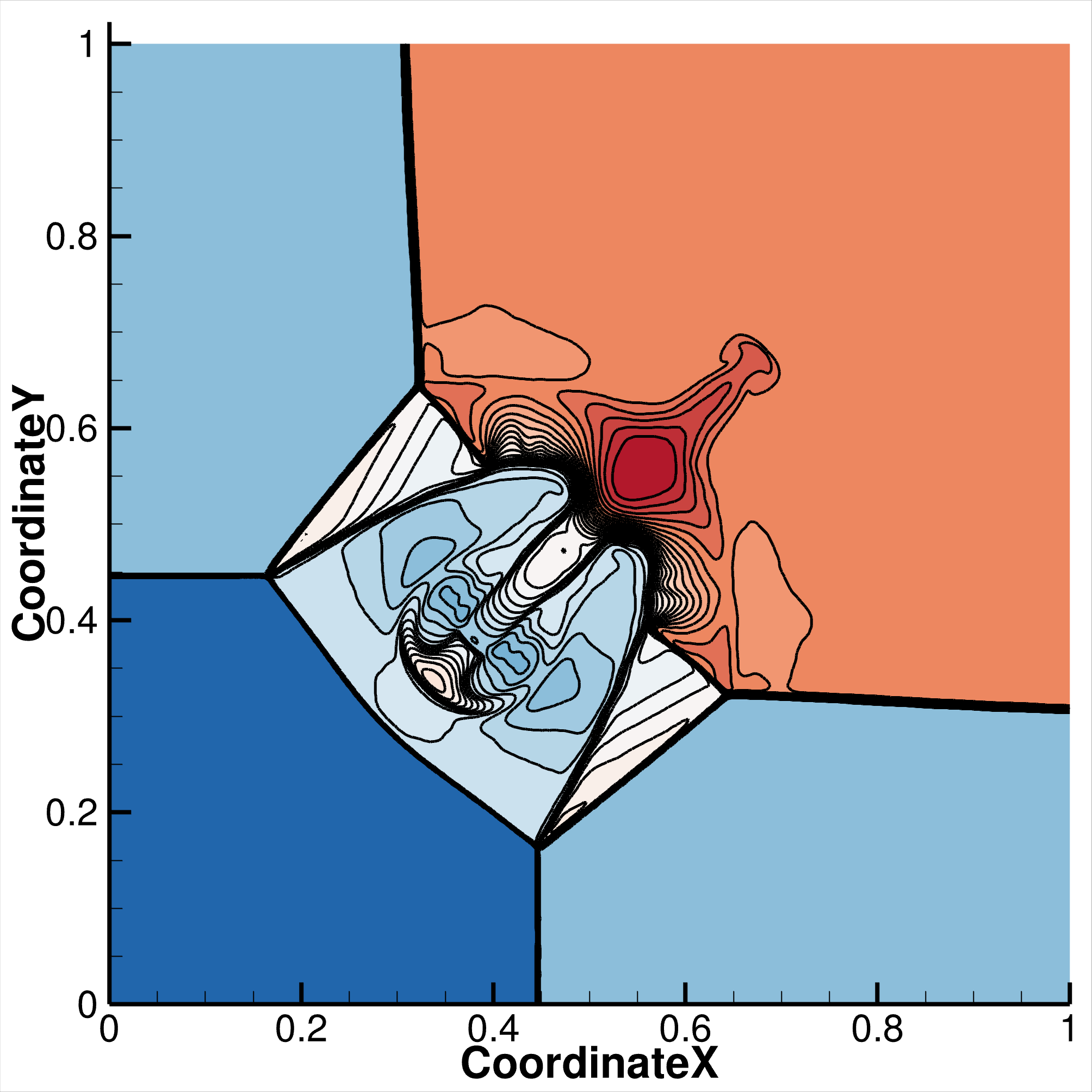}}
  \caption{The density contours (contour levels from 0.2 to 1.7 with increments of 0.05) of two-dimensional shock interaction Riemann problem at time $t=0.6$}\label{fig:riemann1Contor}
\end{figure}

The second test case is the shock interaction problem, a classic test problem for the Euler equations. The initial conditions are set as
\begin{equation*}
  (\rho,V_1,V_2,p)=\begin{cases}
   (1.5,0,0,1.5),        &x>0.7,y>0.7, \\
   (0.5323,1.206,0,0.3), &x<0.7,y>0.7, \\
   (0.138,0,0,0.1),      &x<0.7,y<0.7, \\
   (0.5323,0,1.206,0.3), &x>0.7,y<0.7. \\
  \end{cases}
\end{equation*}
In this case, four initial shock waves interact, resulting in a much more complicated pattern. As analyzed in \cite{lax1998solution}, the initial shock wave $S_{23}^-$ bifurcates at the tip point into a reflected shock wave, a Mach stem, and a slip line. The reflected shock wave interacts with the $S_{12}^-$ shock wave, producing a new shock. The solution of time $t=0.6$ is shown in Figure \ref{fig:riemann1Contor}, which shows that the shock waves interact with each other and result in a much more complicated pattern. The density contours on the stationary mesh and ALE mesh are shown in Figure \ref{fig:riemann1Contor}. 

\begin{figure}[htbp!]
  \centering
  \subfigure[Mesh deformation]{\includegraphics[width=0.4\textwidth]{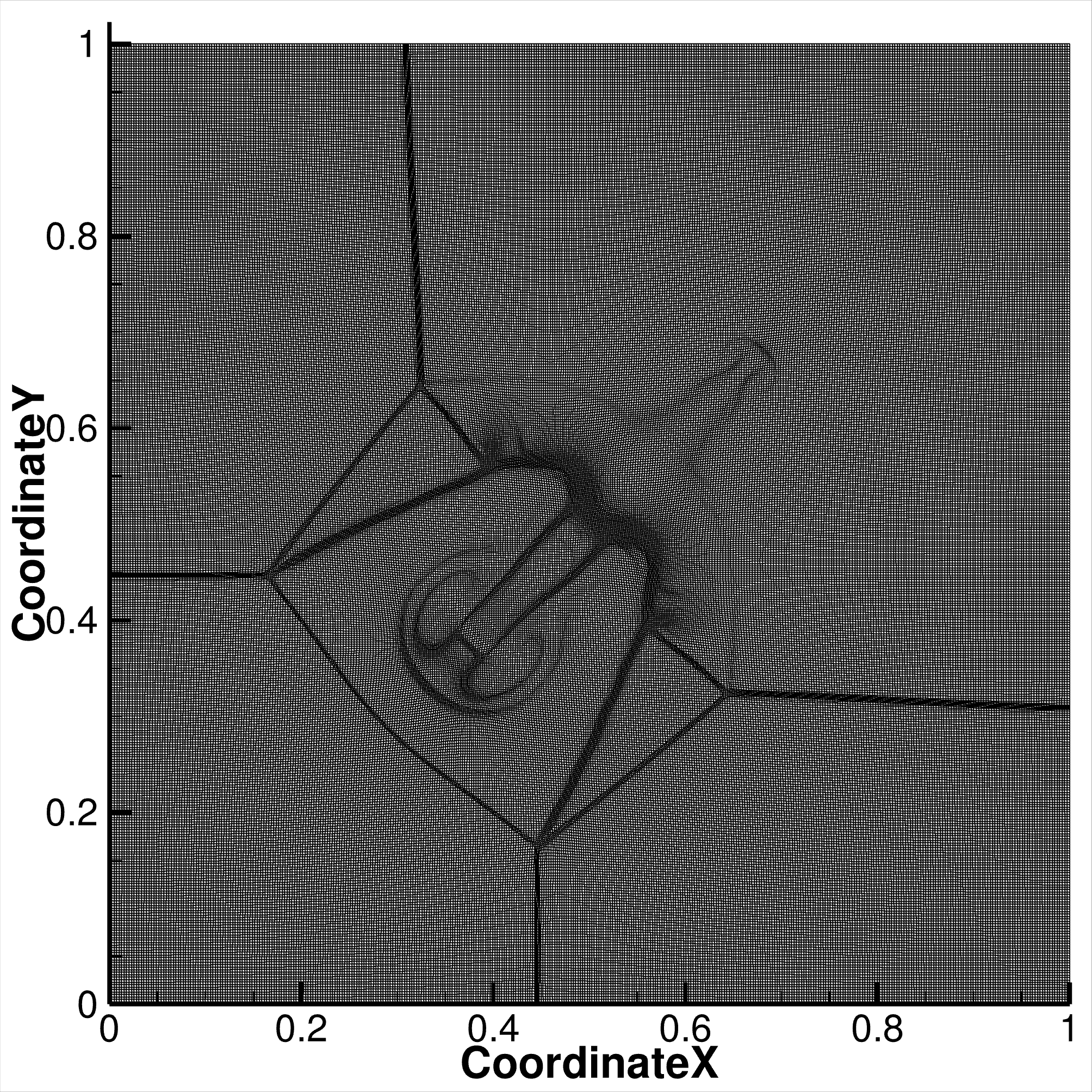}}\quad
  \subfigure[Mesh deformation enlargement]{\includegraphics[width=0.4\textwidth]{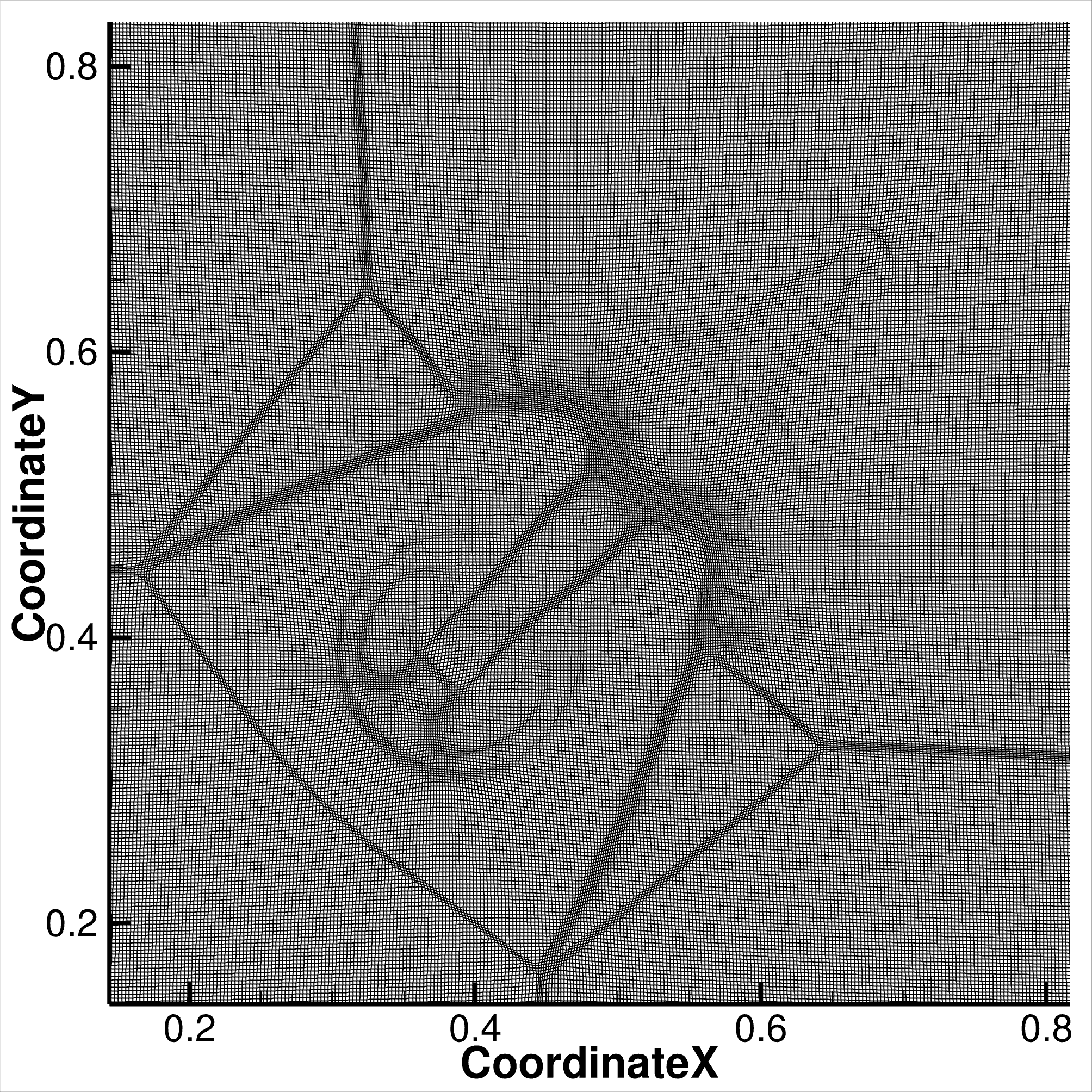}}
  \caption{The mesh distribution of two-dimensional shock interaction Riemann problem at time $t=0.6$}\label{fig:riemann1Mesh}
\end{figure}

A comparison between ALE and stationary-mesh results indicates that the ALE method achieves slightly higher spatial resolution than the stationary-mesh method and demonstrates slight numerical instability at the slip line.
Figure \ref{fig:riemann1Mesh} displays the ALE mesh configuration and local magnification, illustrating adaptive mesh refinement in the shock interaction zone.

\subsection{Spherically symmetric Sod problem}

\begin{figure}[htbp!]
  \centering
  \includegraphics[width=0.4\textwidth]{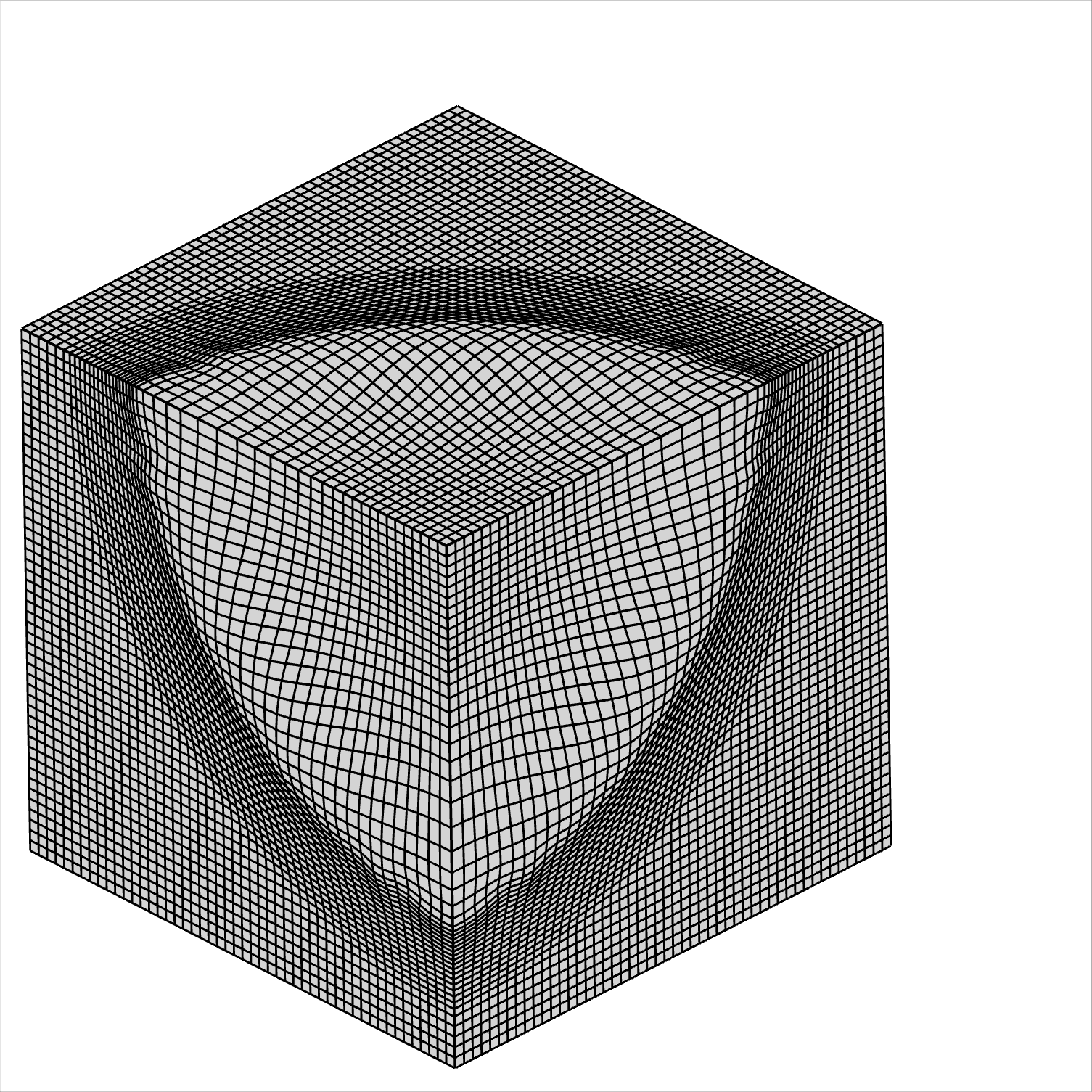}\quad
  \includegraphics[width=0.4\textwidth]{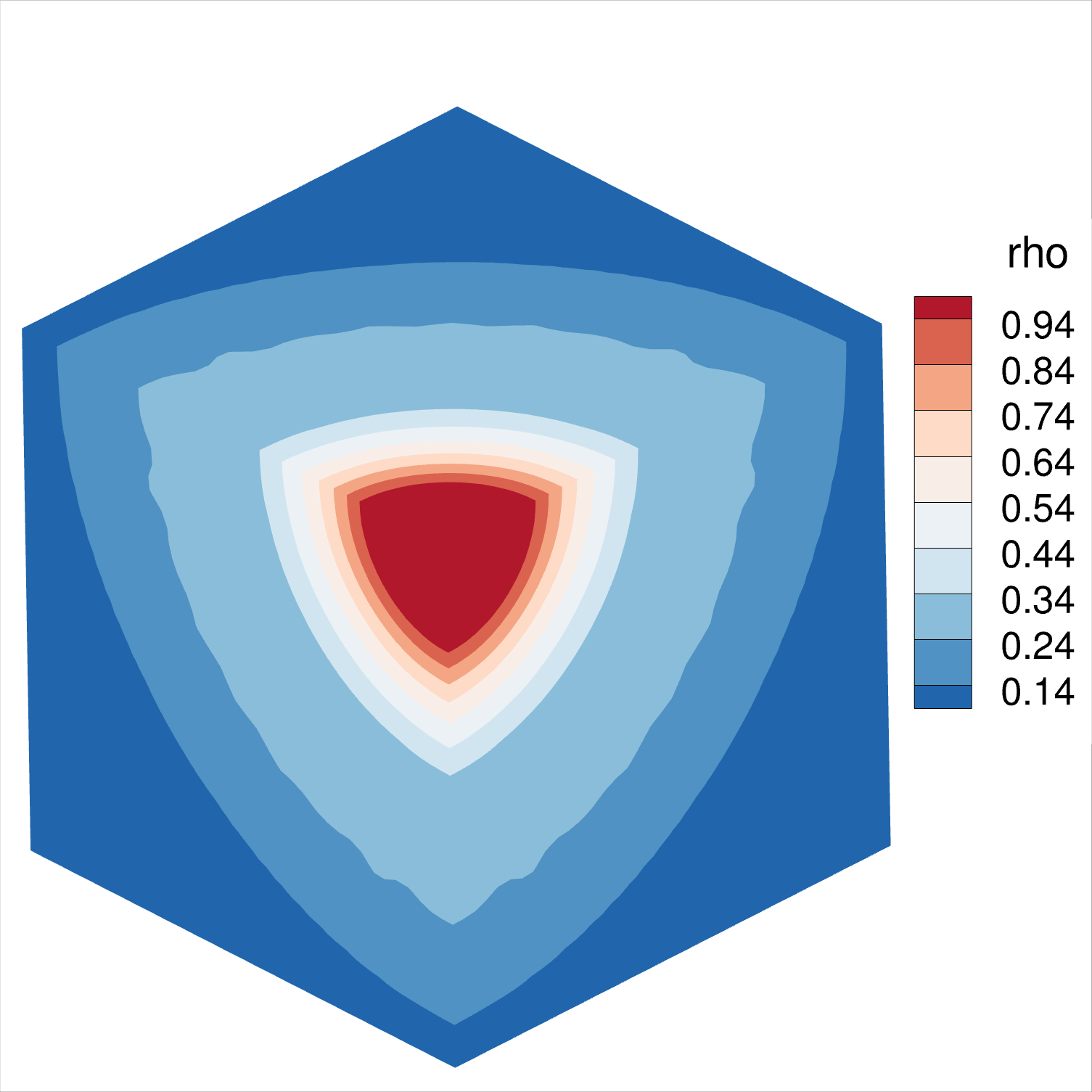}
  \caption{The solution of spherically symmetric Sod problem by Lagrangian velocity}\label{fig:sodlang}
\end{figure}

\begin{figure}[htbp!]
  \centering
  \includegraphics[width=0.4\textwidth]{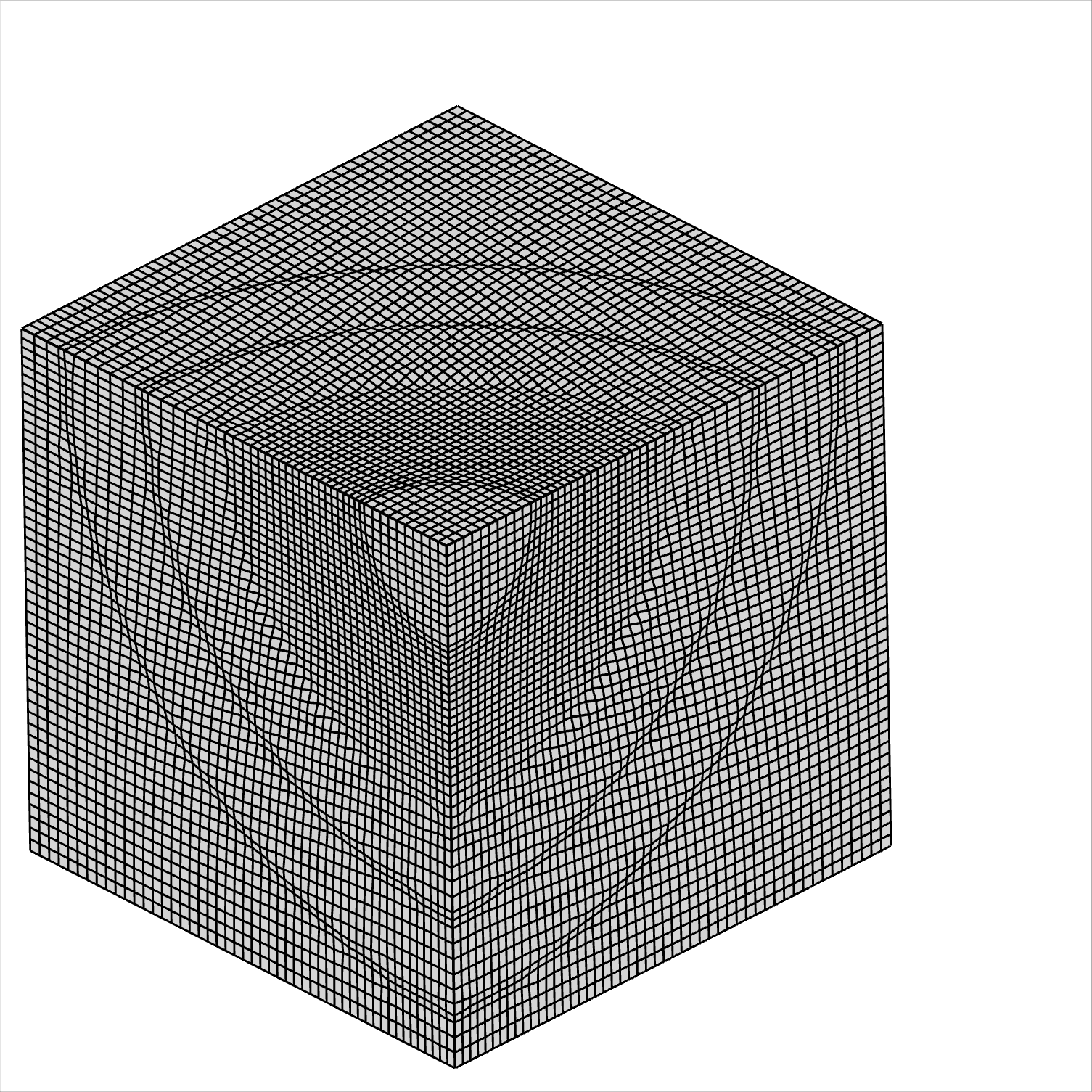}\quad
  \includegraphics[width=0.4\textwidth]{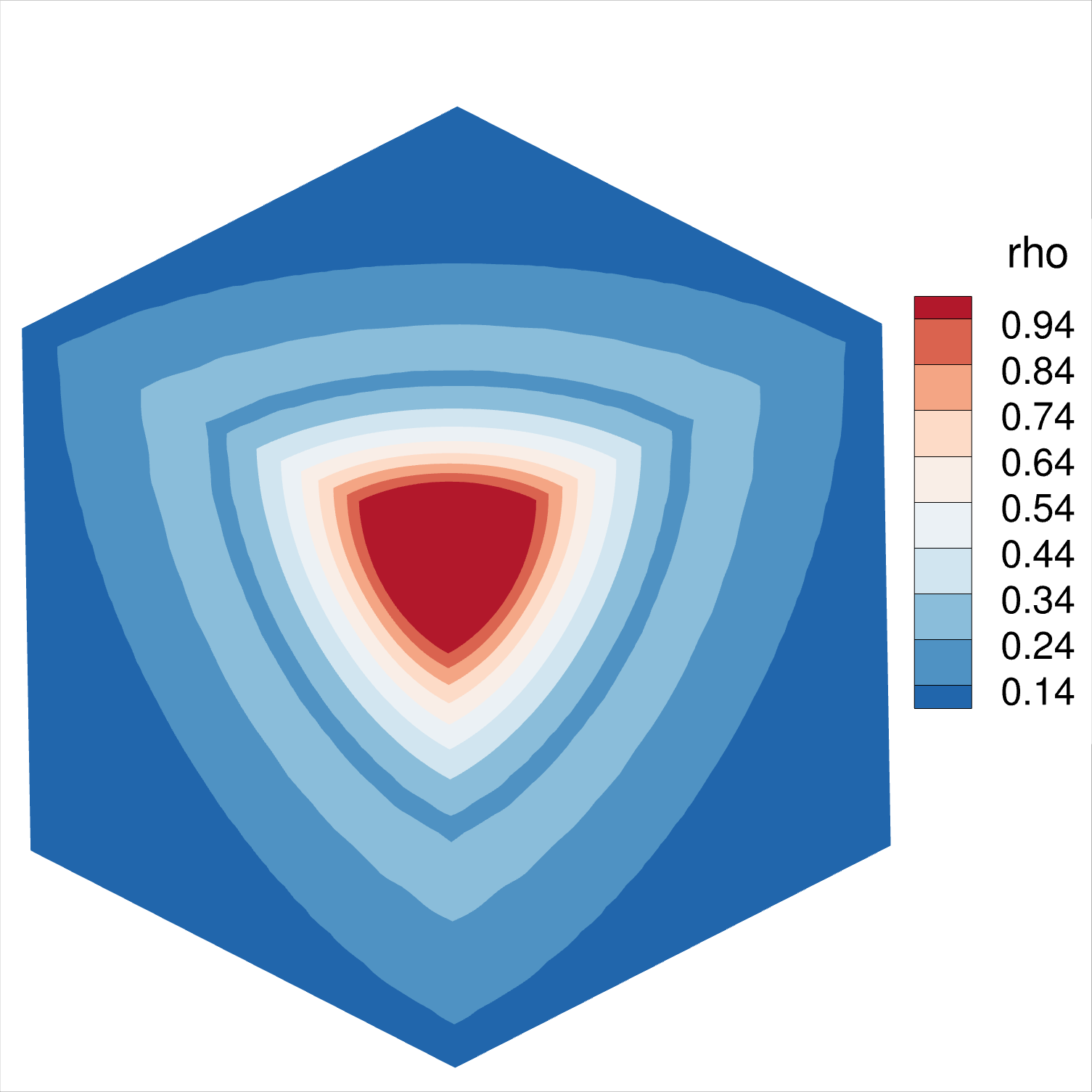}
  \caption{The solution of spherically symmetric Sod problem by variational approach}\label{fig:sodvar}
\end{figure}

The spherically symmetric Sod problem is a standard test problem for the Riemann problem. The initial conditions are set as
\begin{equation*}
  (\rho,V_1,V_2,V_3,p)=\begin{cases}
   (1,0,0,0,1),        &r<0.5 \\
   (0.125,0,0,0,0.1), &r>0.5 \\
  \end{cases}
\end{equation*}
 The computational domain is $(x, y,z) \in[0,1]\times[0,1]\times[0,1]$, and the mesh number is $50\times50\times50$. The symmetric boundary condition is imposed on the plane with $x = 0$, $y = 0$, and $z = 0$, and the non-reflection boundary condition is imposed on the plane with $x = 1$, $y = 1$, and $z = 1$. The exact solution of the spherically symmetric problem can be given by the following one-dimensional system with geometric source terms
\begin{equation*}
  \frac{\partial }{\partial t}\left(\begin{matrix}
    \rho\\ \rho U\\ \rho E 
  \end{matrix}\right) + \frac{\partial }{\partial r}\left( \begin{matrix}
    \rho U\\ \rho U^2+p\\ \rho UE+pU\\
  \end{matrix}\right) =-\frac{d-1}{r} \left( \begin{matrix}
    \rho U\\ \rho U^2\\ \rho UE+pU\\
  \end{matrix}\right),
  \end{equation*}
  where the radial direction is denoted by $r$, $U$ is the radial velocity, $d$ is the number of space dimensions. Both the variational approach and the Lagrangian velocity method are used for mesh deformation. For the variational approach, the coefficient is set to $\alpha=1.0$. For the Lagrangian velocity method, an additional mesh-smoothing process is applied every 20 time steps with a relaxation coefficient $\omega=0.5$. The CFL number is set to 0.3 for this case. The mesh distribution and the density contours at time $t=0.25$ are shown in Figure \ref{fig:sodvar} and Figure \ref{fig:sodlang}.

\begin{figure}[htbp!]
  \centering
  \includegraphics[width=0.4\textwidth]{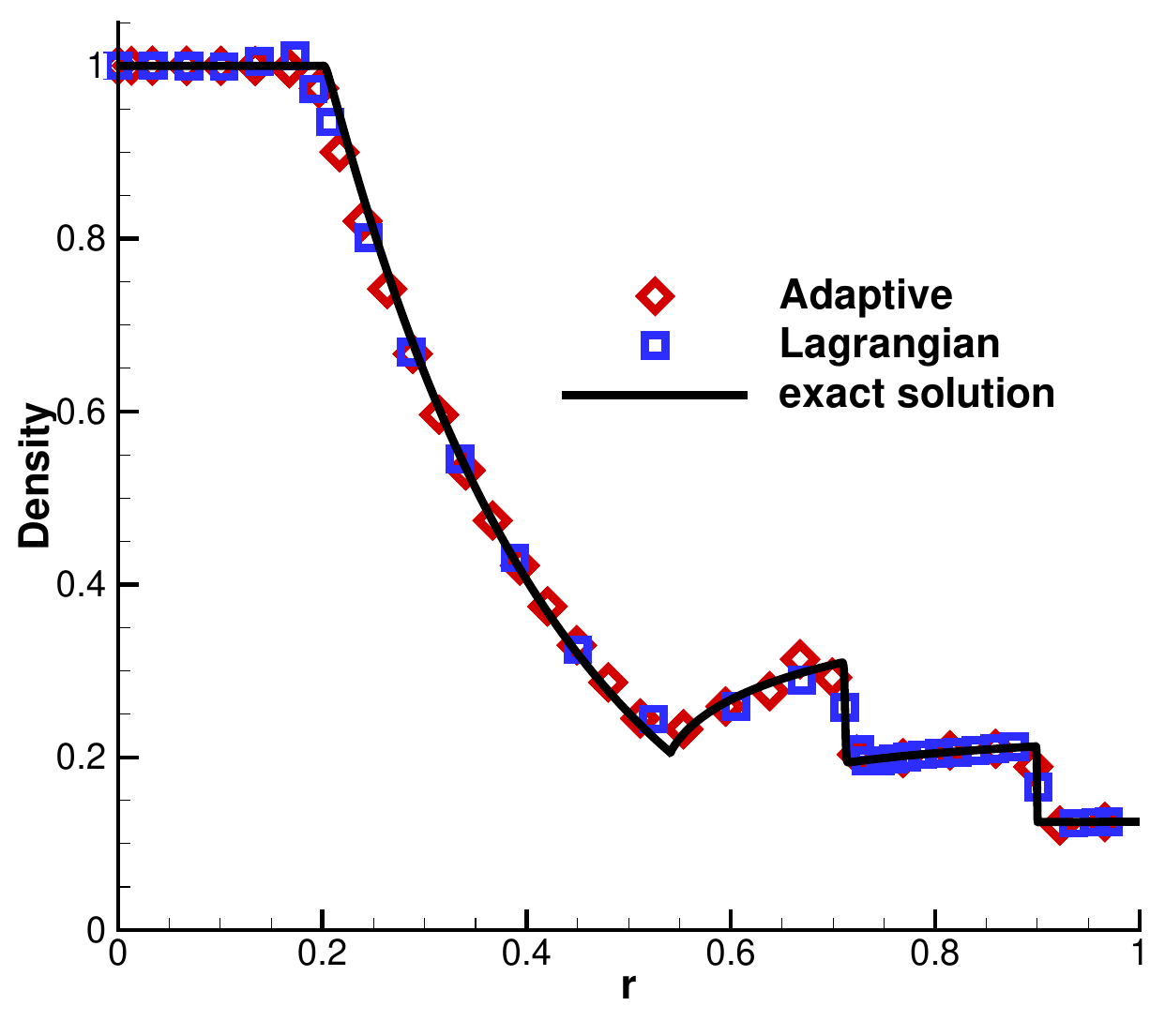}\quad
  \includegraphics[width=0.4\textwidth]{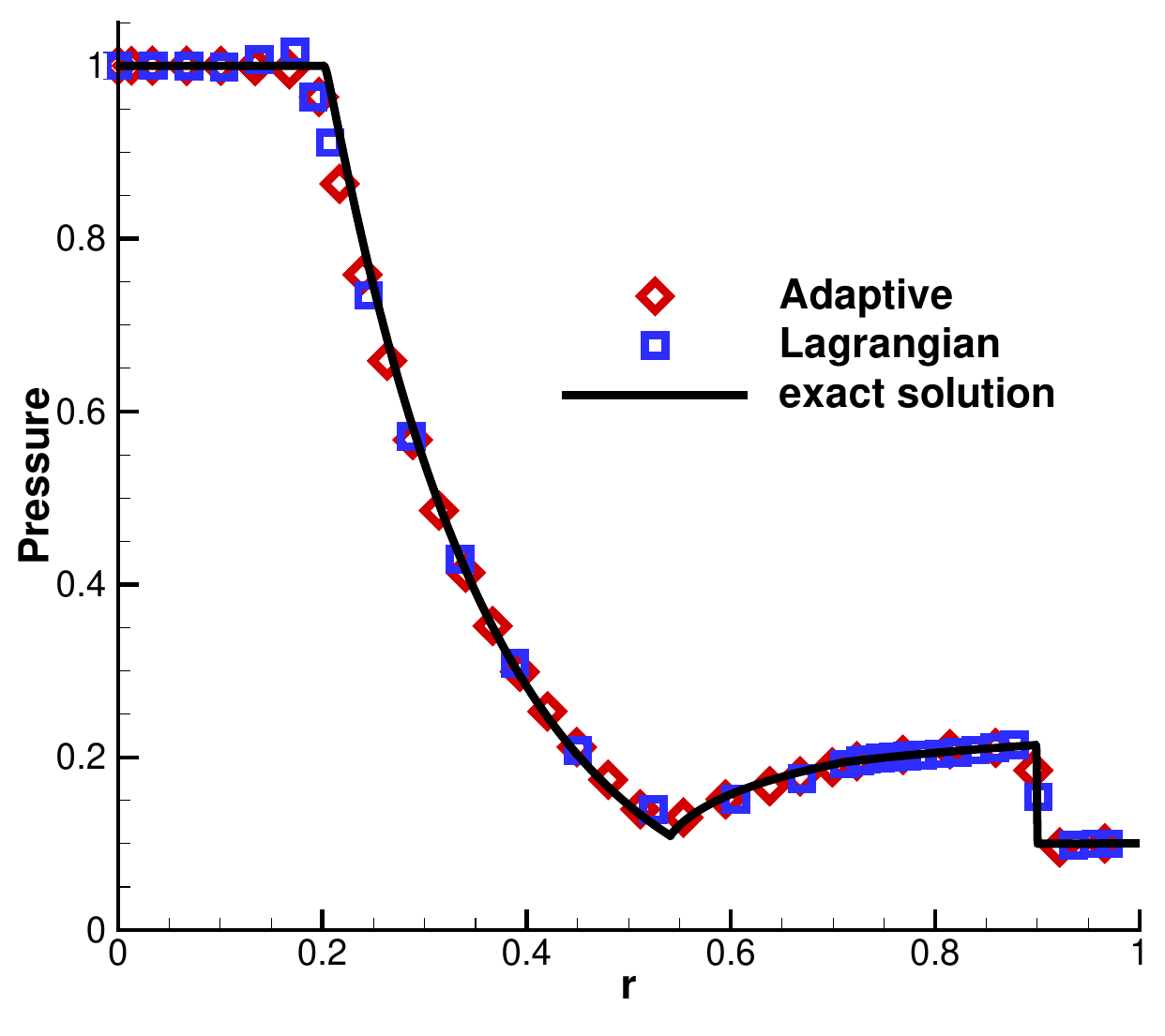}
  \caption{The density and pressure distribution along the diagonal line $x=y=z$}\label{fig:sodrho}
\end{figure}

For the Lagrangian velocity method, the mesh distribution is shown in Figure \ref{fig:sodlang}. The mesh is refined in the region between the shock and the contact discontinuity. For the variational approach, the mesh distribution is shown in Figure \ref{fig:sodvar}, where the mesh is refined at the shock, contact discontinuity, and rarefaction wave. The density and pressure distribution along the diagonal line $x=y=z$ is shown in Figure \ref{fig:sodrho}. The solution agrees well with the exact solution.
\subsection{Sedov Problem}

\begin{figure}[hbt!]
  \centering
   \subfigure[Mesh distribution]{ \includegraphics[width=0.4\textwidth]{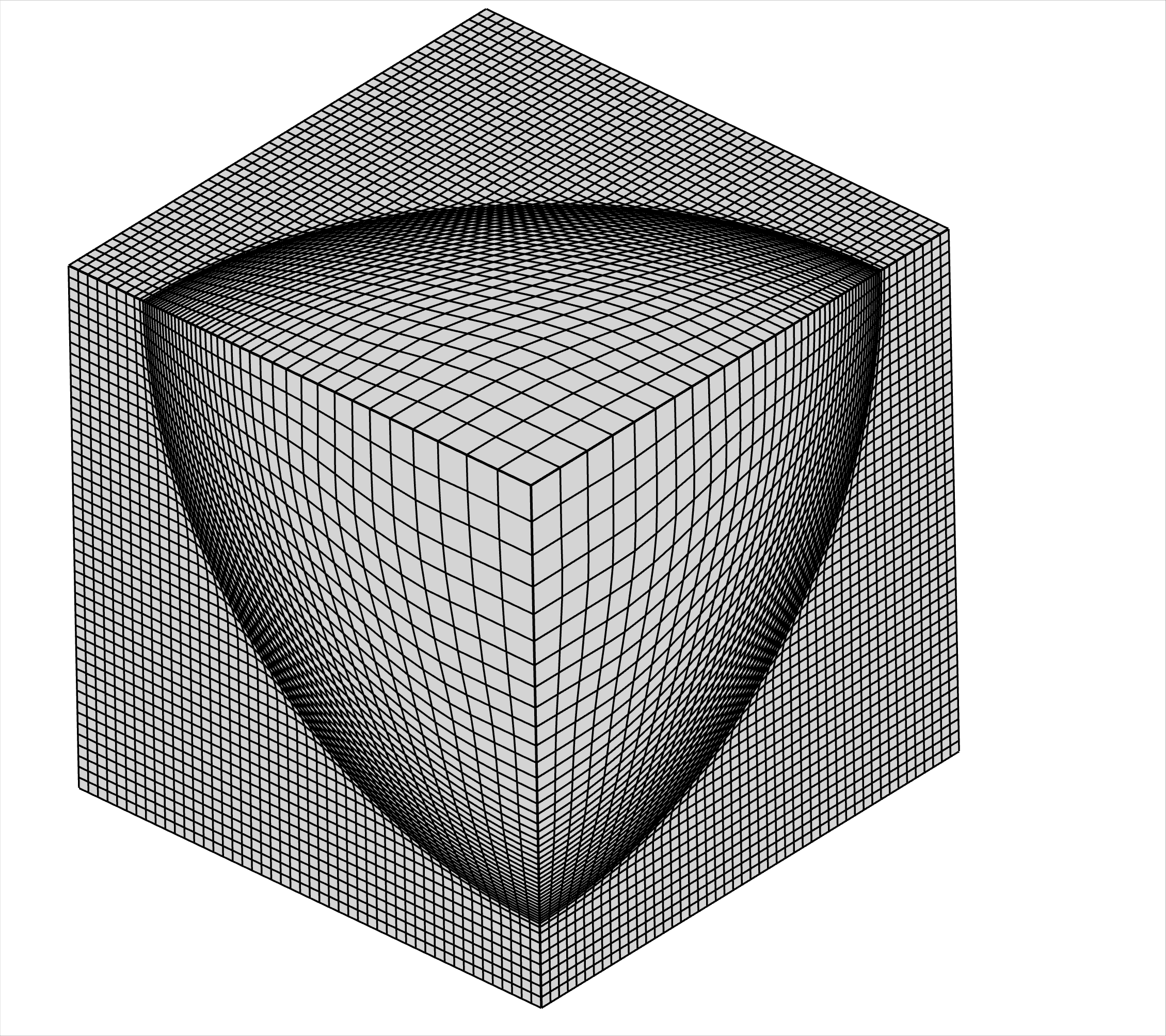}}\quad
  \subfigure[Density contours]{ \includegraphics[width=0.4\textwidth]{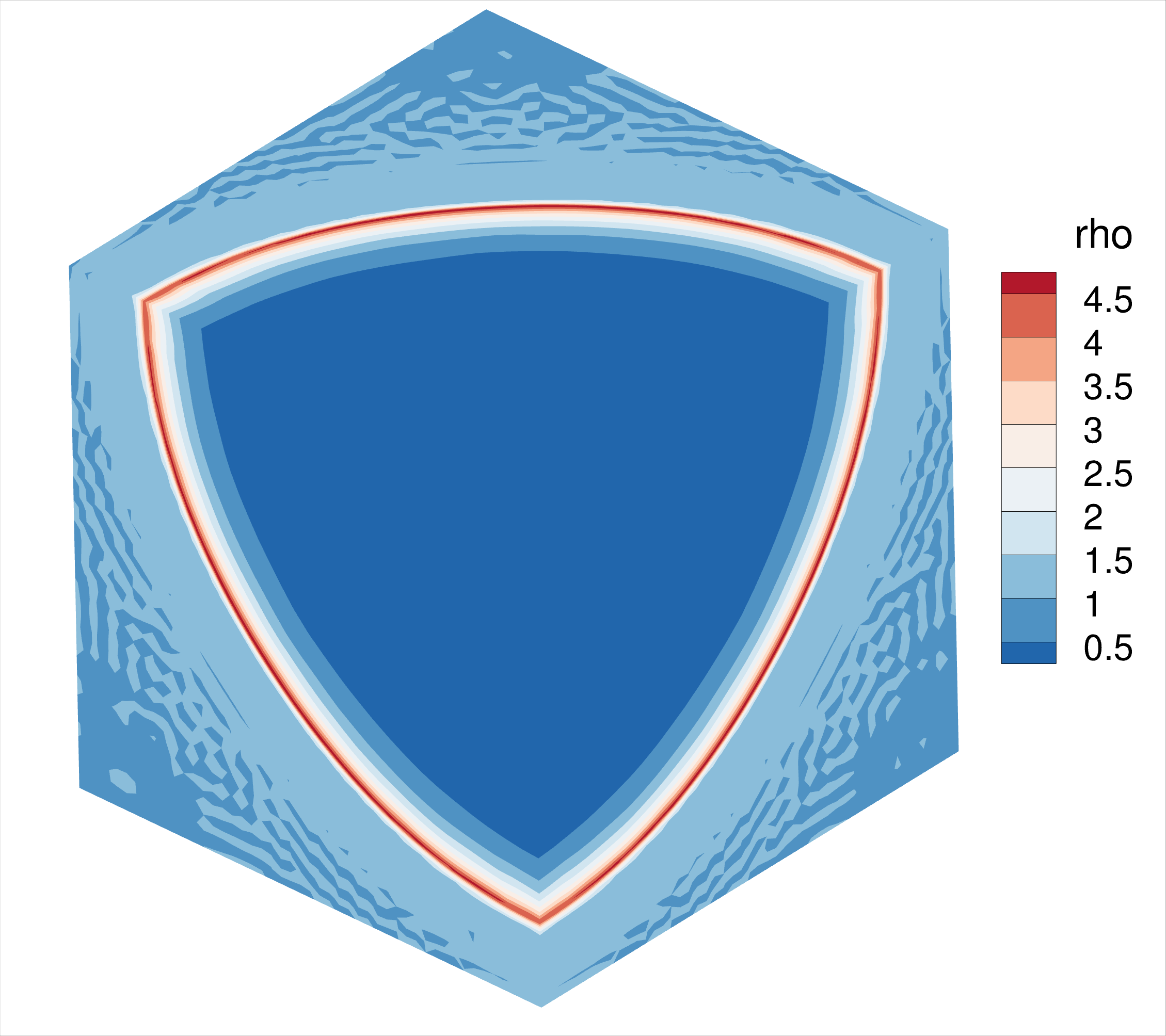}}
  \caption{Sedov problem: the mesh distribution and density contours at $t=1$}\label{sedov_contour}
\end{figure}

\begin{figure}[hbt!]
  \centering
   \subfigure[Density]{ \includegraphics[width=0.4\textwidth]{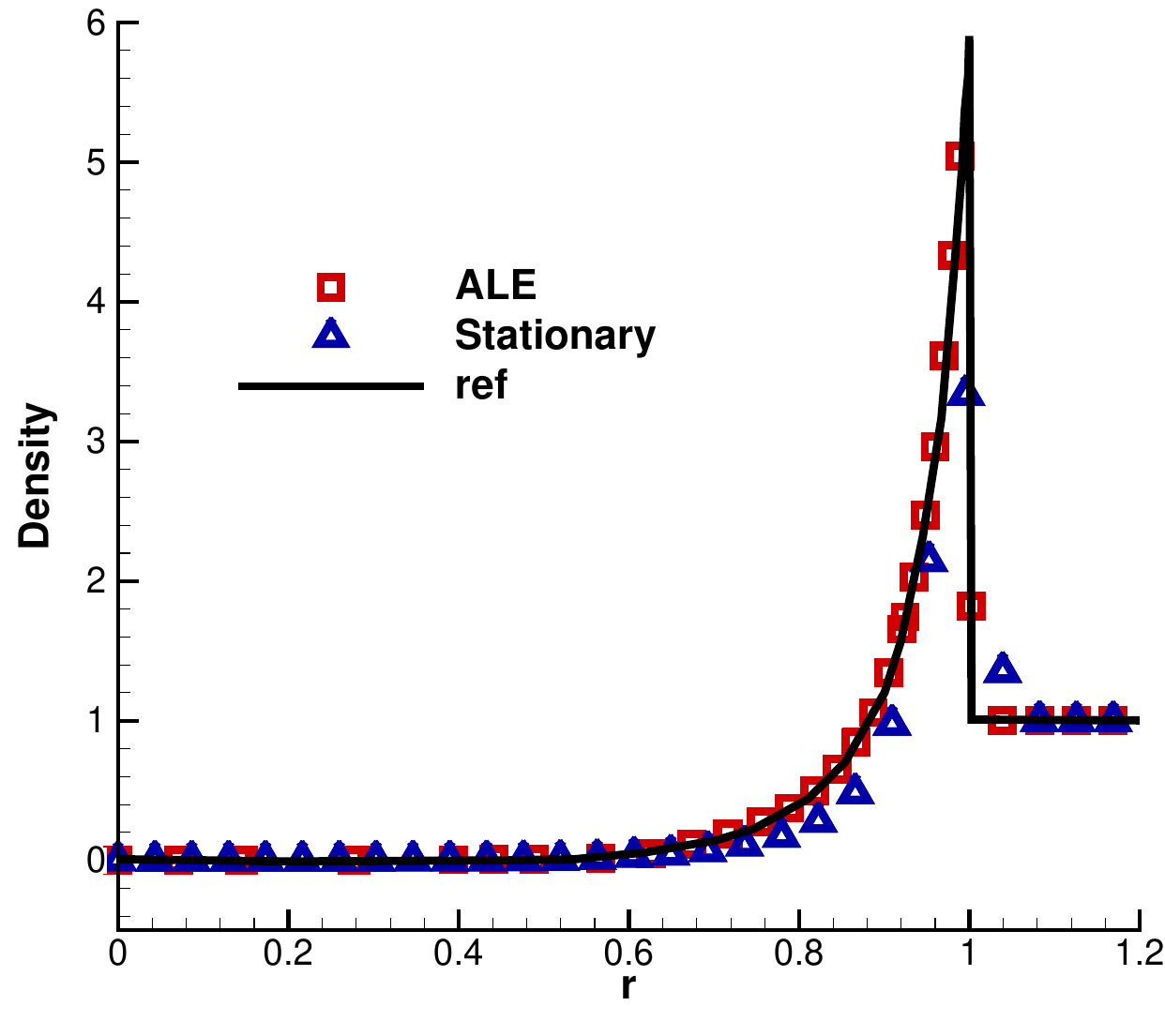}}\quad
    \subfigure[Pressure]{ \includegraphics[width=0.4\textwidth]{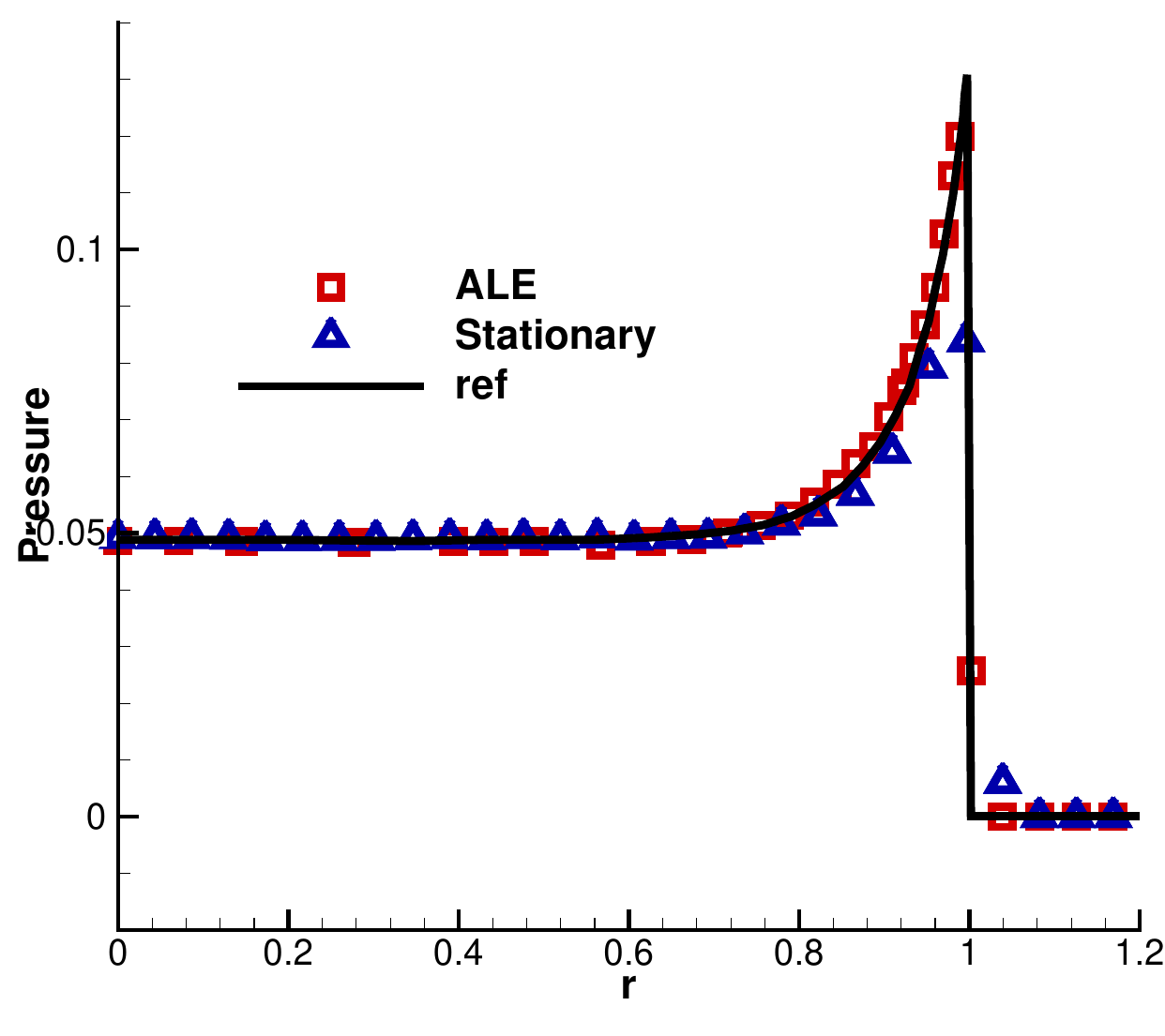}}
  \caption[]{Sedov problem: the density and pressure distribution along diagonal line}\label{sedov_line}
\end{figure}

The Sedov blast wave is a standard problem in hydrodynamics that models a blast wave emanating from a point source of energy. It is a benchmark problem for both the Lagrangian velocity method and the ALE method. The computational domain is $[0,1.2]^3$, and a uniform grid with $h=1/40$ is used as the initial mesh in this test case. The fluid is modeled by the ideal gas equation of state with $\gamma=1.4$. The initial condition is a uniform static field with density $\rho=1$ and pressure $p=1e-6$, except for the cell containing the origin. In this cell, the initial pressure is defined by $p=(\gamma-1)\epsilon_0/V$, where $\epsilon_0= 0.106384$ is the total amount of released energy and $V$ is the cell volume. Asymmetric boundary condition is enforced at the plane $x=0$, $y=0$ and $z=0$, while non-reflecting boundary conditions are used at the plane $x=1.2$, $y=1.2$, and $z=1.2$. Mesh movement is accomplished using Lagrangian velocity, and a smoothing process is applied every five time-steps with a relaxation coefficient of $\omega=0.5$. A small CFL number of 0.01 is used to avoid the instability caused by the initial singularity at the origin. After ten steps, the simulation transitions to a normal CFL number of 0.3. Figure \ref{sedov_contour} displays the mesh distribution and density contour at time $t=1$, which reveals that the mesh is refined adaptively around the shock.
 Figure \ref{sedov_line} displays the density and pressure distribution along the diagonal from $(0,0,0)$ to $(1.2,1.2,1.2)$, as well as the solutions of a uniform stationary mesh.
The results show that solutions from the ALE method agree more closely with the exact solution than those from the static mesh method. The ALE method provides solutions at extreme points that are closer to the exact solution than those from the uniform mesh.

\subsection{Noh Problem}

\begin{figure}[hbt!]
  \centering
    \subfigure[Mesh distribution with mesh $36^3$]{  \includegraphics[width=5cm]{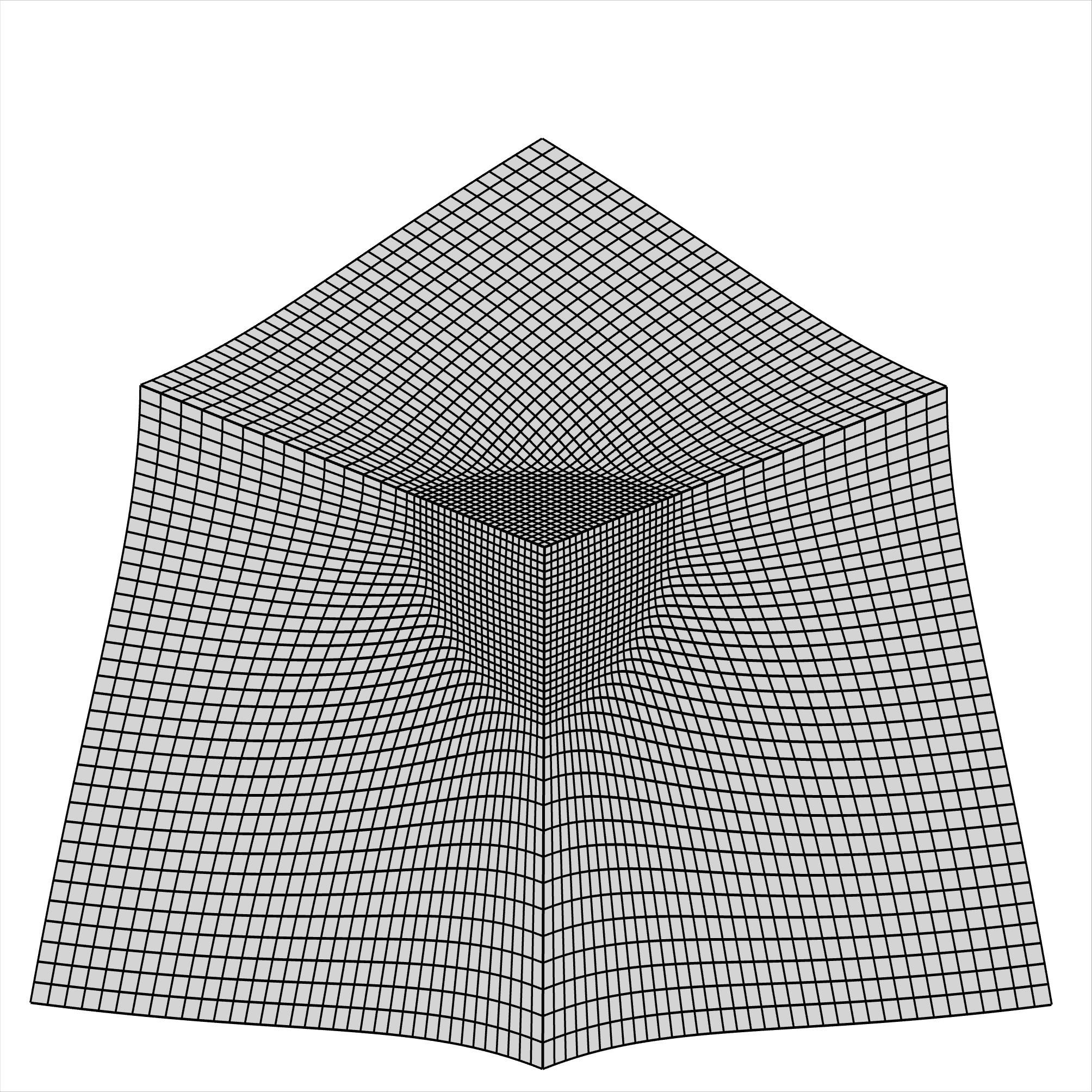}}\quad
   \subfigure[Mesh distribution with mesh $72^3$]{  \includegraphics[width=5cm]{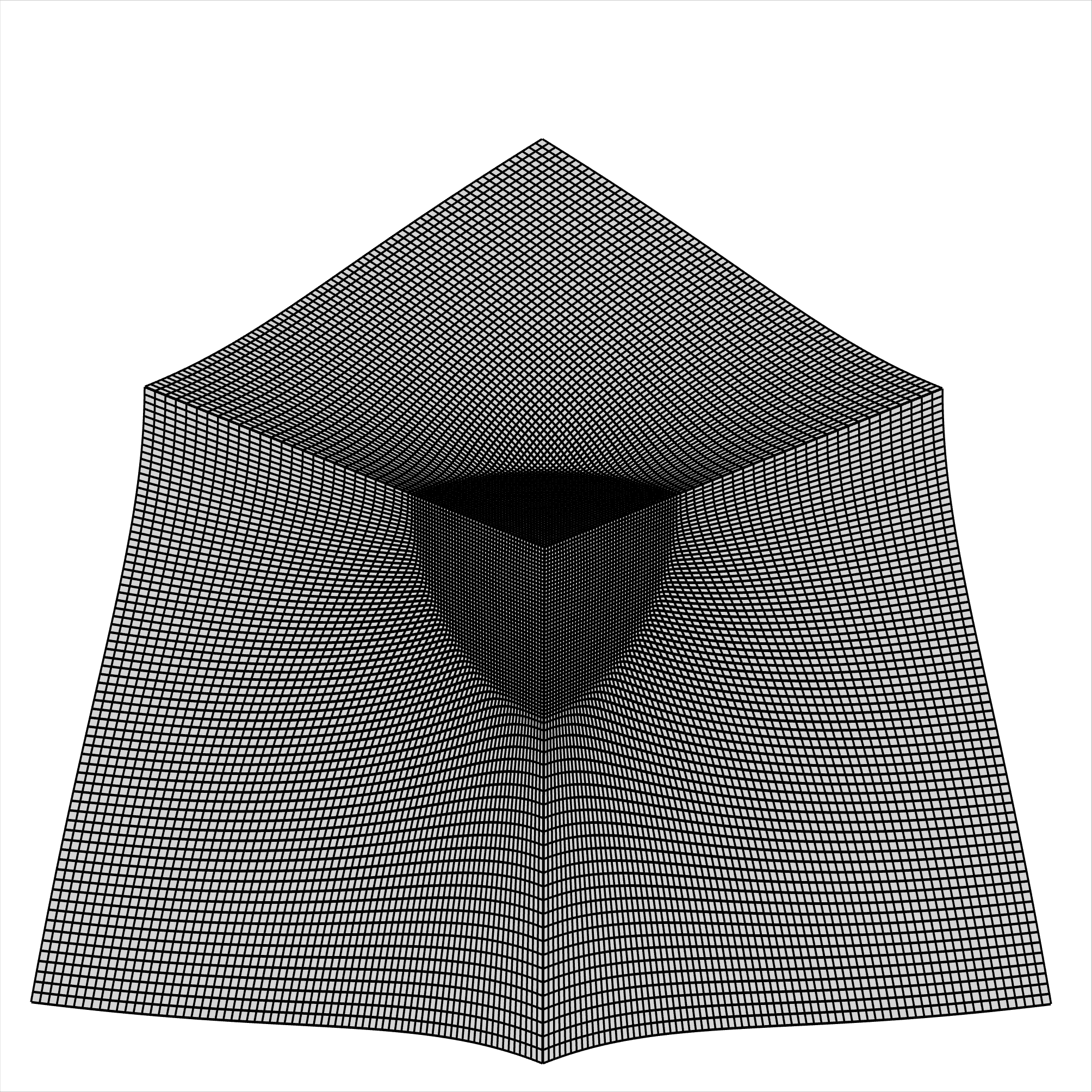}}
  \caption[]{The mesh distribution with mesh $36^3$ and $72^3$ used in Noh Problem}\label{noh_final_mesh}
\end{figure}

This case is a typical test case for verifying ALE code. A three-dimensional domain with dimensions $[0,1.2]\times[0,1.2]\times[0,1.2]$ is considered. Two meshes with mesh numbers $36^3$ and $72^3$ are used for this test case. The initial conditions are set to uniform density $\rho=1$ and internal energy $e=p/(\gamma-1)=1\times10^{-4}$, with $\gamma=5/3$. The initial velocity is given by $\mathbf{V}=(-x/r,-y/r,-z/r)$, where $r$ is the distance from the origin. An axially symmetric shock is generated at the origin and propagates at a constant speed. At time $t=0.6$, the shock is located at $r=0.2$. An exact solution of density \cite{NOH1987} is
\begin{equation*}
  \rho=\begin{cases}
    64,&r<0.2 \\ (1+t/r)^2,&r>0.2
  \end{cases}.
\end{equation*}
An asymmetric boundary condition is enforced at the lines $x=0$, $y=0$, and $z=0$, while a non-reflective boundary condition is used at all other boundaries. Mesh movement is achieved using Lagrangian velocity, and a smoothing process is applied every 5 time-steps with a relaxation coefficient $\omega=0.8$. The CFL number is set to 0.05 for this case. The final meshes with $36^3$ and $72^3$ are shown in Figure \ref{noh_final_mesh}, and it is observed that the mesh is concentrated in the post-shock region.

\begin{figure}[hbt!]
  \centering
    \subfigure[Density contour with mesh $36^3$]{ \includegraphics[width=5cm]{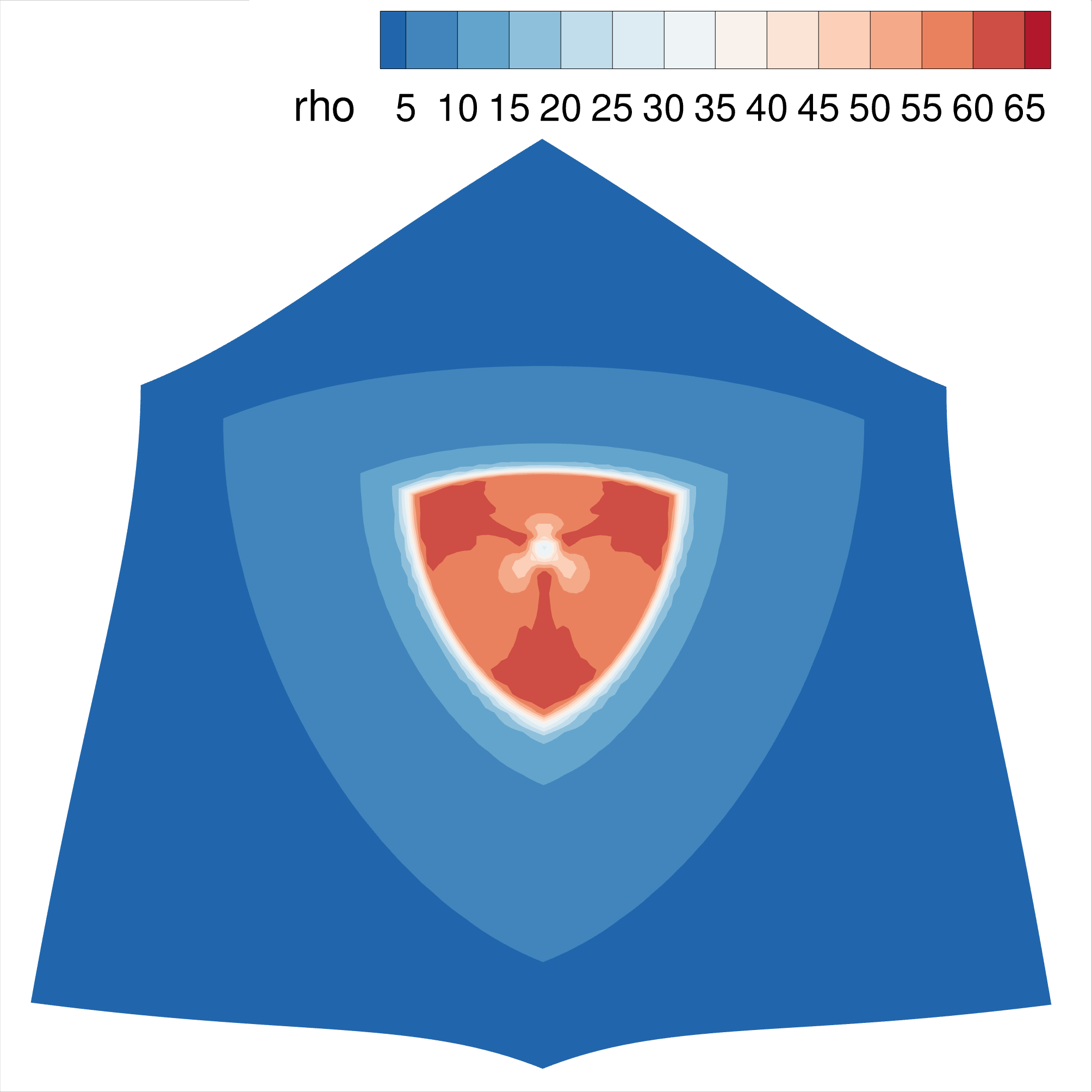}}\quad
   \subfigure[Density contour with mesh $72^3$]{  \includegraphics[width=5cm]{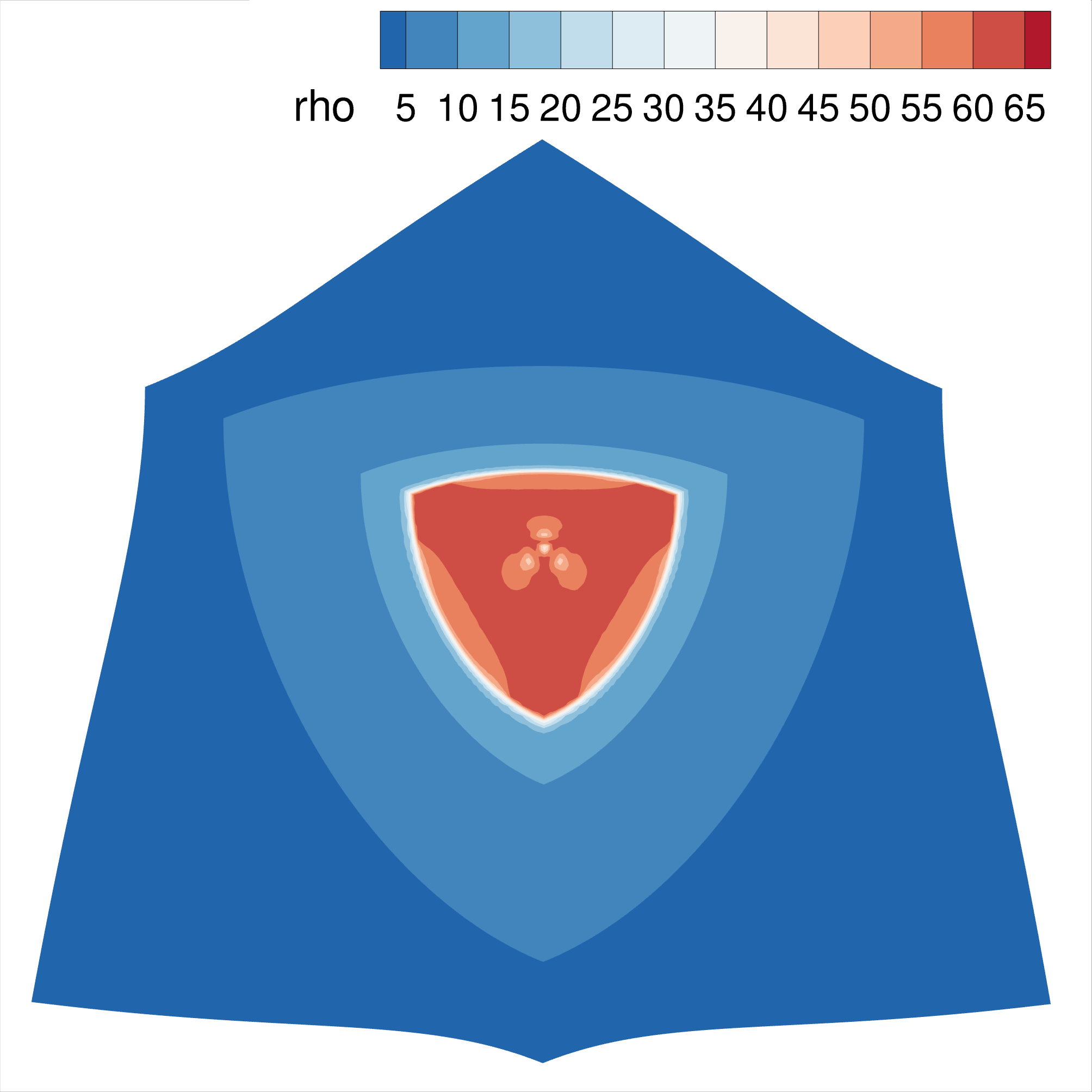}}
  \caption[]{The density contour with mesh $36^3$ and $72^3$ used in Noh Problem}\label{noh_dens_contour}
\end{figure}

\begin{figure}[hbt!]
  \centering
 \subfigure[The density distribution with respect to $r$ with mesh $36^3$]{ \includegraphics[width=0.4\textwidth]{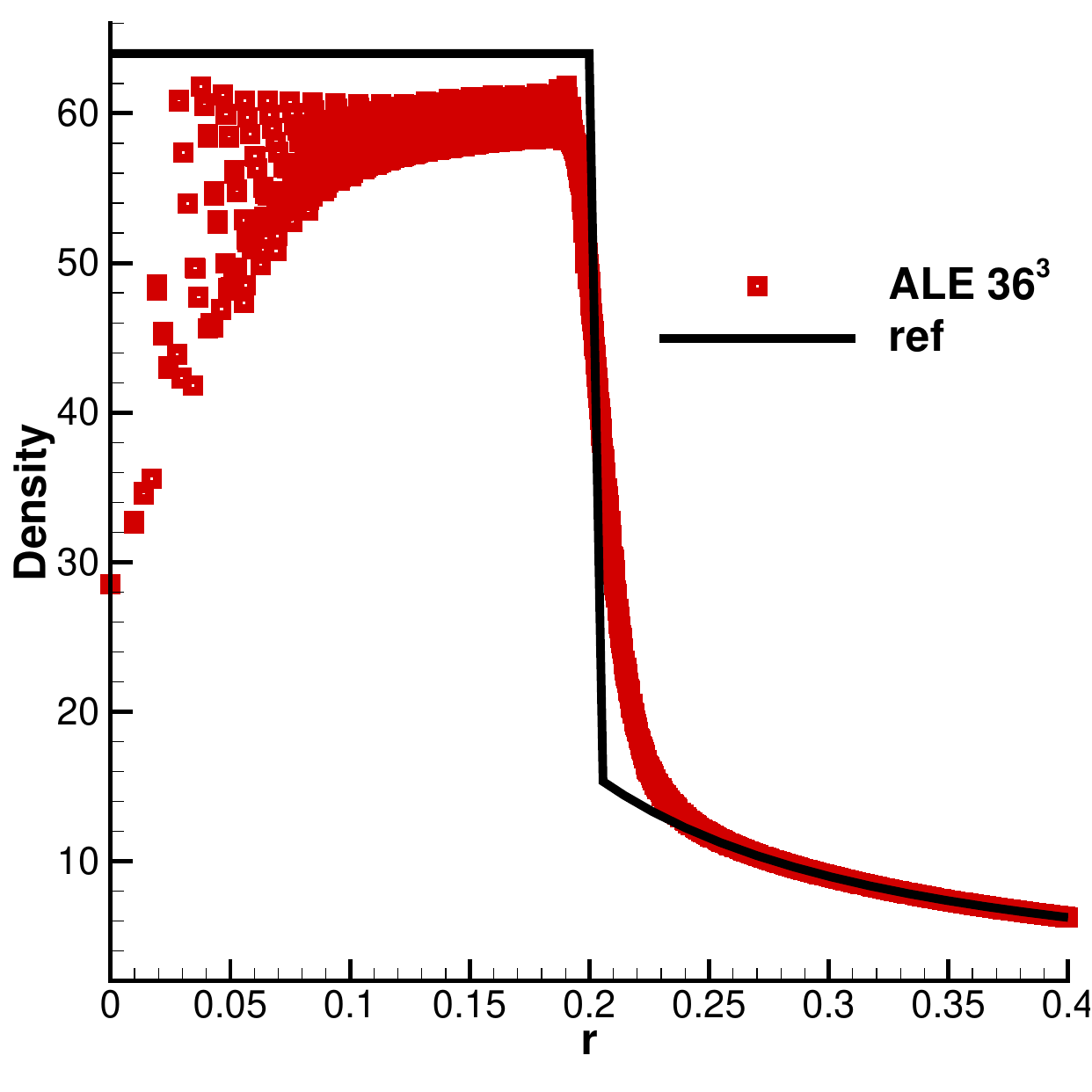}}\quad
 \subfigure[The density distribution with respect to $r$ with mesh $72^3$]{ \includegraphics[width=0.4\textwidth]{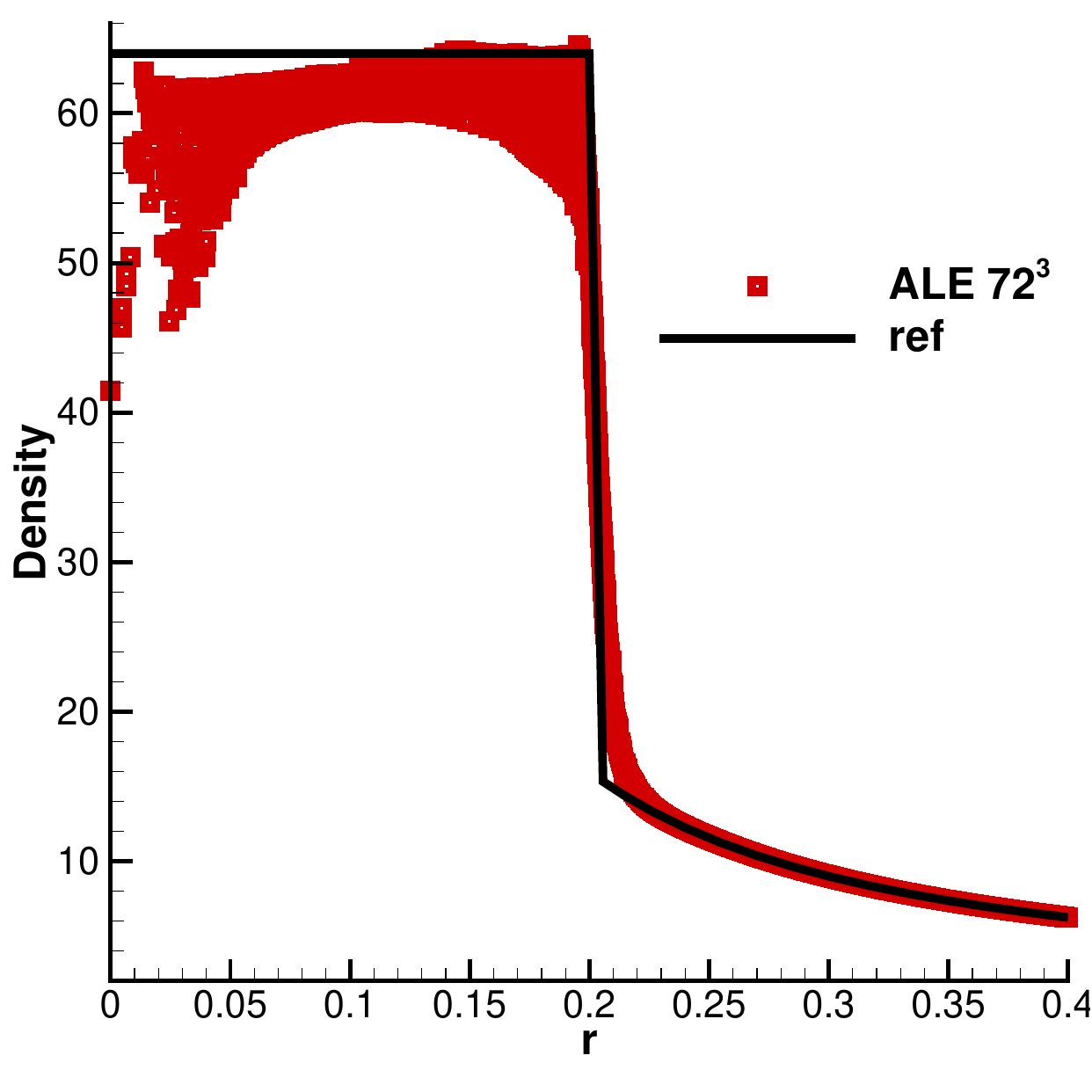}}
 \caption[]{The density distributions with respect to $r$ of Noh problem at $t=0.6$}\label{noh_dens_dist}
\end{figure}

The density contours at time $t=0.6$ are shown in Figure \ref{noh_dens_contour}, 
and the distributions with respect to $r$ in the entire domain are presented in Figure \ref{noh_dens_dist}. Good agreements between the numerical and exact solutions have been observed. A comparison between the coarse and fine meshes shows that the fine mesh can obtain a more accurate solution and less numerical oscillation.

\subsection{Saltzmann Problem}

\begin{figure}[hbt!]
  \centering
   \includegraphics[width=0.8\textwidth]{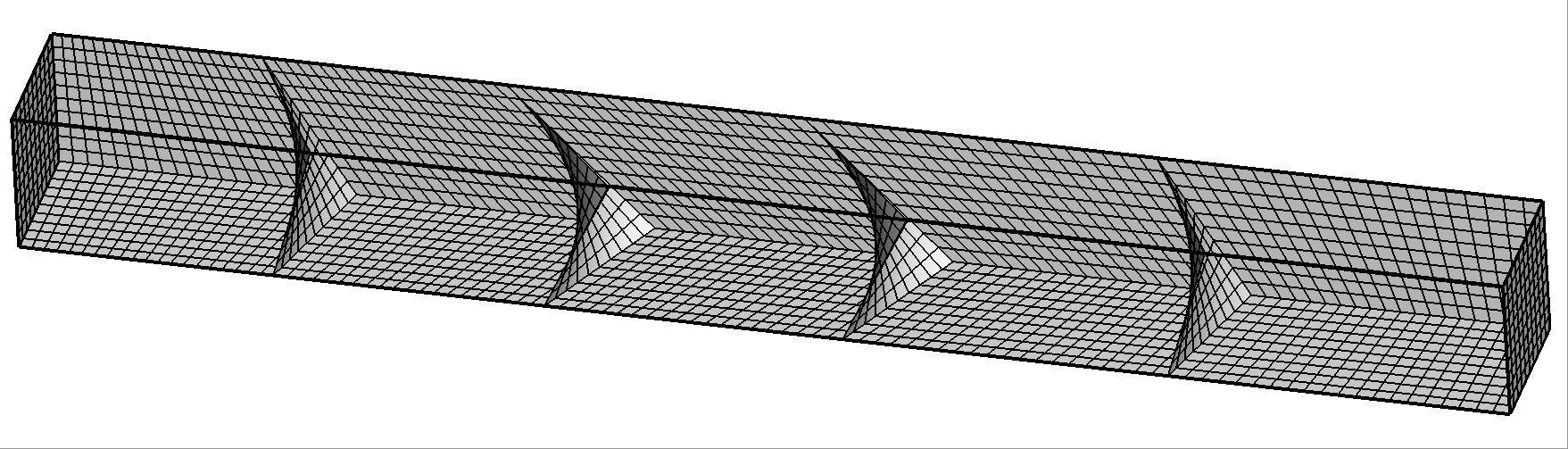}
  \caption[]{The initial mesh of Saltzmann problem} \label{Saltzmann_mesh}
\end{figure}

The Saltzmann problem is a test case that simulates fluid flow in a cylinder with a piston. It examines the motion of a planar shock on a skewed Cartesian grid. The computational domain is $[0,1]\times[0,0.1]\times[0,0.1]$. The grid is uniform with $h=0.01$ in three directions. A mesh distortion of the $x$-coordinate is defined by
 \begin{equation*}
  \Delta x = (0.5y+z-15yz)\sin(\pi x).
 \end{equation*}
 The corresponding initial mesh is shown in Figure \ref{Saltzmann_mesh}, as referenced above.
  Initially, an ideal monatomic gas with $\rho=1,e=10^{-4},\gamma=5/3$ fills the box. The wall on the left moves at a constant speed of 1 to the right. It acts as a piston. The other boundaries act as inviscid reflecting walls. As the gas is compressed, a shock is generated. It propagates to the right and moves faster than the piston. Mesh movement uses Lagrangian velocity. A smoothing process is applied every 20 time steps with a relaxation coefficient of $\omega=0.8$.
  
\begin{figure}[hbt!]
  \centering
  \subfigure[Density]{ \includegraphics[width=0.4\textwidth]{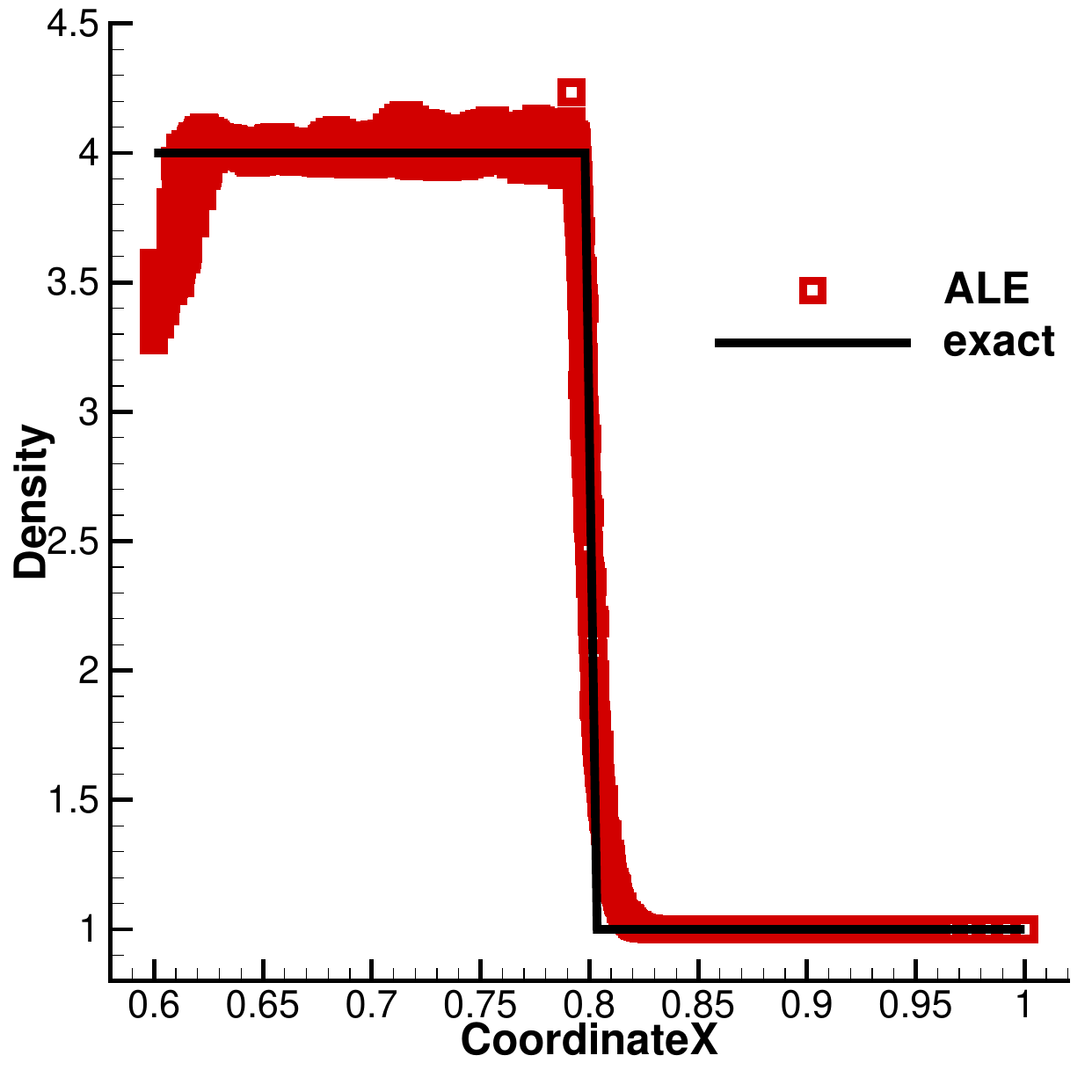}}\quad
  \subfigure[Pressure]{ \includegraphics[width=0.4\textwidth]{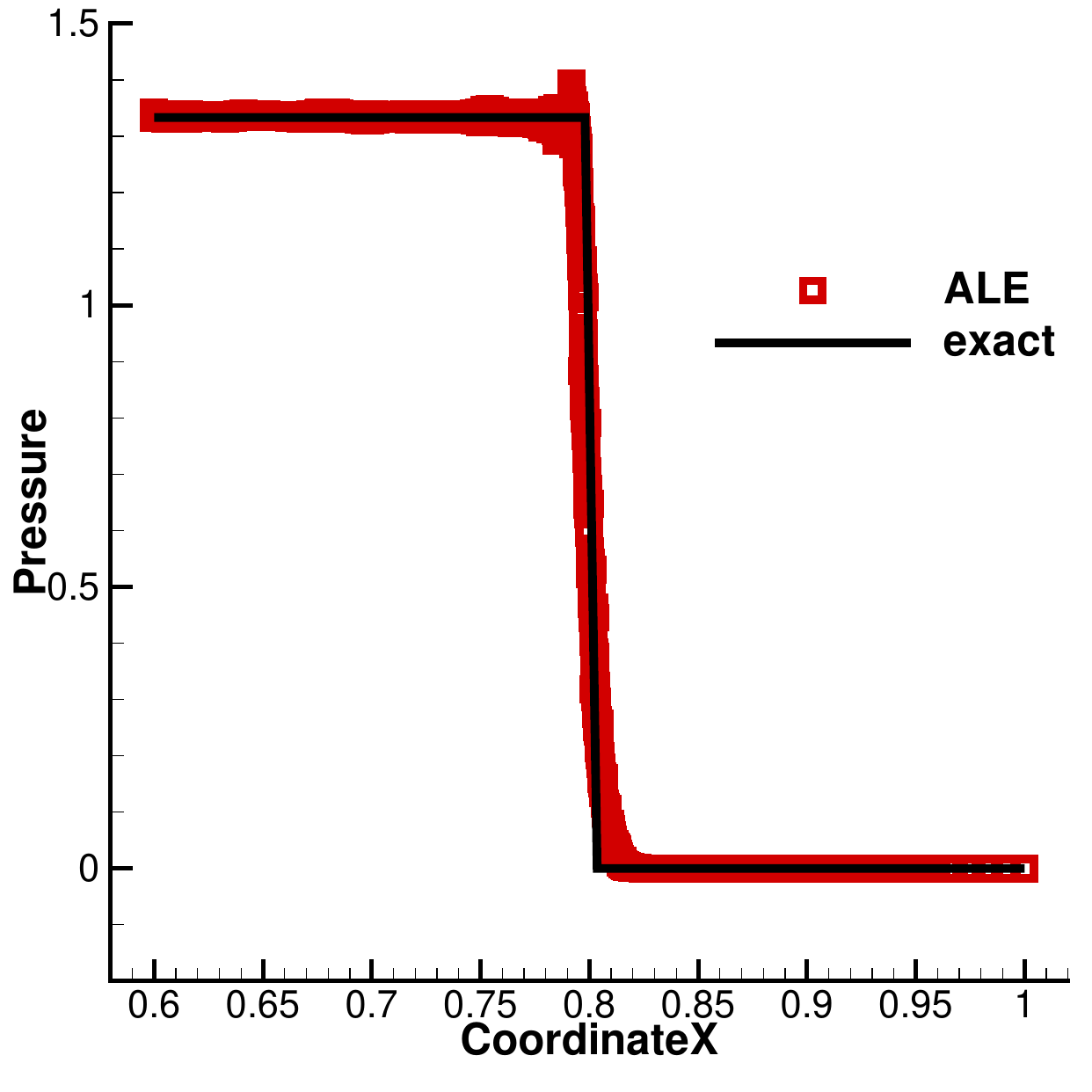}}
  \caption[]{Saltzmann problem: the density and pressure distribution at $t=0.6$}\label{Saltzmann_line}
\end{figure}

\begin{figure}[hbt!]
  \centering
   \subfigure[$t=0.6$]{ \includegraphics[width=0.4\textwidth]{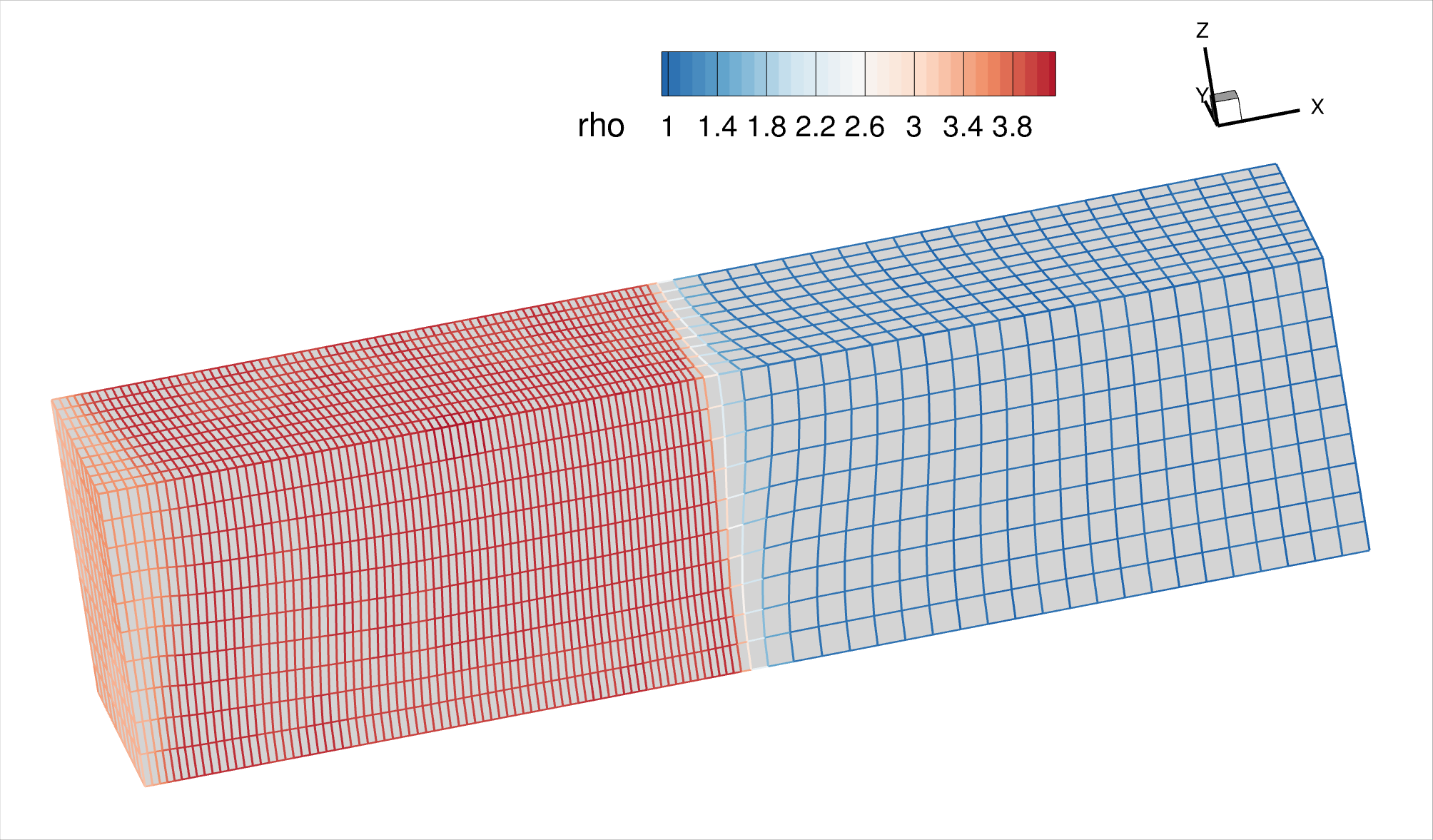}}\quad
   \subfigure[$t=0.85$]{ \includegraphics[width=0.4\textwidth]{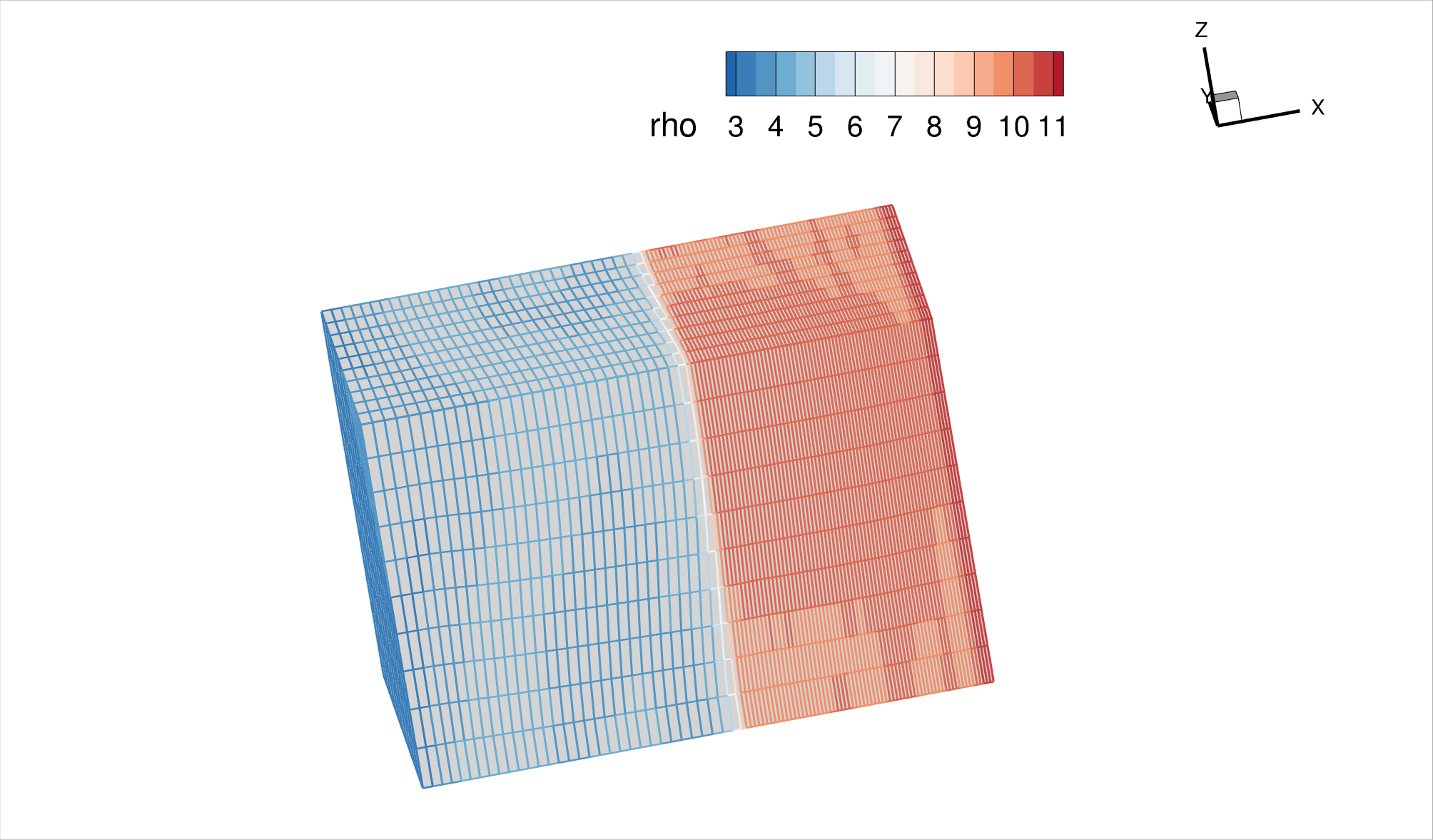}}
  \caption[]{Saltzmann problem: mesh and density distribution at different time}\label{Saltzmann_time}
\end{figure}

  Figure \ref{Saltzmann_line} shows the entire solution of the domain at $t=0.6$, indicating the solution tends towards a one-dimensional result.
The drop in density may be due to an increase in entropy at the wall. At $t=0.6$, the shock is located at $x=0.8$, and the post-shock solutions are $\rho=4$ and $p=1.333$. At time $t=0.85$, the shock bounces back due to the wall. Figure \ref{Saltzmann_time} demonstrates that the Lagrangian velocity method is capable of handling shock motion issues, as it shows the mesh distribution at times $t=0.6$ and $t=0.85$.

\section{Conclusion}
In this paper, a high-order compact gas-kinetic scheme is developed within an arbitrary Lagrangian-Eulerian (ALE) framework on a three-dimensional structured mesh. 
The gas-kinetic formulation provides time-accurate solutions at cell interfaces, enabling the use of a single-stage third-order flux evaluation, unlike Runge–Kutta methods that require multiple sub-steps.
The compact GKS is used in this paper, which uses a small stencil to achieve high-order accuracy. The compact reconstruction is based on the GKS, which provides the evolution of both the cell-averaged flow variables and their gradients. 
Based on the compact stencil, a memory-reduced reconstruction is used to reduce computational cost \cite{liu2024memory}.
Test results show a 2.4$\times$ to 3.0$\times$ speedup in computational efficiency over the baseline method.
The generalized ENO (GENO) method was also used in this paper to achieve high-order accuracy and robustness by introducing a path function that combines a fourth-order polynomial with the second-order ENO reconstruction.
The numerical test cases from the low speed viscous flow to the high speed shock wave demonstrate the robustness and accuracy of our method.

 \section*{Acknowledgements}
This work was supported by the National Key R\&D Program of China (Grant No. 2022YFA1004500), the National Natural Science Foundation of China (Nos. 12172316, 92371107, 12302378 and 92371201), the Hong Kong Research Grant Council (Nos. 16301222, and 16208324) and the Funding of National Key Laboratory of Computational Physics, and the Natural Science Basic Research Plan in Shaanxi Province of China (No. 2025SYS-SYSZD-070).

% %% The Appendices part is started with the command \appendix;
% %% appendix sections are then done as normal sections
% %% \appendix

% %% \section{}
% %% \label{}

\bibliographystyle{elsarticle-num}
\bibliography{mybibfile}
%\printbibliography
\end{document}